\documentclass[a4paper,12pt]{article}

\usepackage{subfloat}
\usepackage{amsmath,amssymb,amsfonts}
\allowdisplaybreaks
\usepackage{graphicx}
\usepackage{color}
\usepackage{caption}
\usepackage{subcaption}
\makeatletter
\@addtoreset{equation}{section}
\renewcommand{\theequation}{\thesection.\@arabic\c@equation}
\makeatother
\usepackage{hyperref}
\usepackage{cite}
\usepackage{caption}
\definecolor{red}{rgb}{1,0,0}
\definecolor{green}{rgb}{0,1,0}
\definecolor{blue}{rgb}{0,0,1}
\definecolor{darkblue}{rgb}{0,0,0.5}
\definecolor{lightblue}{rgb}{.5,.5,1}
\definecolor{lightgray}{gray}{.87}
\definecolor{Dark}{gray}{.20}
\definecolor{pink}{rgb}{.95,0.82,0.92}
\definecolor{yellow}{rgb}{1,1,0}
\definecolor{lightyellow}{rgb}{1,1,.5}
\definecolor{purple}{rgb}{0.7,0,0.85}
\definecolor{darkgreen}{rgb}{0,0.25,0}
\definecolor{darkpurple}{rgb}{0.4,0,0.4}
\definecolor{orange}{rgb}{0.8,0.2,0.2}
\def \be {\begin{equation}}
\def \ee {\end{equation}}
\def \bea {\begin{align}}
\def \eea {\end{align}}
\def \nn {\nonumber}

\def \rr {\raise.35ex\hbox{\small $\prime$}\kern-.17em{\mbox{\large $\imath$}}}
\def \del {\partial}
\def \dels {\partial\kern-.5em / \kern.5em}
\def \As {{A\kern-.5em / \kern.5em}}
\def \Ds {D\kern-.7em / \kern.5em}

\def \a {\alpha}

\def \b {\beta}
\def \dag {\dagger}

\def \d {\delta}

\def \lam {\lambda}
\def \Lam {\Lambda}

\def \om {\omega}
\def \Om {\Omega}

\def \th {\theta}
\def \Th {\Theta}

\newcommand{\detail}[1]{}

\newcommand{\hide}[1]{}

\newcommand{\explanation}[1]{}

\pdfoutput=1

\begin{document}

\pagestyle{plain}


\begin{titlepage}
\vspace*{-10mm}   
\baselineskip 10pt   
\begin{flushright}   
\begin{tabular}{r} 
\end{tabular}   
\end{flushright}   
\baselineskip 24pt   
\vglue 10mm

\begin{center}

\noindent
\textbf{\LARGE
UV And IR Effects On Hawking Radiation
\vskip0.3em
}
\vskip20mm
\baselineskip 20pt

\renewcommand{\thefootnote}{\fnsymbol{footnote}}

{\large
Pei-Ming~Ho
\footnote[1]{pmho@phys.ntu.edu.tw},
Hikaru~Kawai
\footnote[2]{hikarukawai@phys.ntu.edu.tw}
}

\renewcommand{\thefootnote}{\arabic{footnote}}

\vskip5mm

{\it
Department of Physics and Center for Theoretical Physics, \\
National Taiwan University, Taipei 106, Taiwan,
R.O.C. 
\\
Physics Division, National Center for Theoretical Sciences, \\
Taipei 106, Taiwan,
R.O.C. 
}

\vskip 25mm
\begin{abstract}

We study the time-dependence of Hawking radiation
for a black hole in the Unruh vacuum,
and find that it is not robust against certain UV and IR effects.
If there is a UV cutoff at the Planck scale,
Hawking radiation is turned off after the scrambling time.
In the absence of a UV cutoff,
Hawking radiation is sensitive to the IR cutoff
through a UV/IR connection
due to higher-derivative interactions in the effective theory.
Furthermore,
higher-derivative interactions with the background
contribute to a large amplitude of particle creation
that changes Hawking radiation.
This unexpected large effect is related to
a peculiar feature of the Hawking particle wave packets.

\end{abstract}
\end{center}

\end{titlepage}

\pagestyle{plain}

\baselineskip 18pt

\setcounter{page}{1}
\setcounter{footnote}{0}
\setcounter{section}{0}


\newpage


\section{Introduction}\label{introduction}


It is well-known that Hawking radiation suffers from 
the trans-Planckian problem \cite{trans-Planckian-1}.
Nevertheless, 
a common opinion is that low-energy effective theories
suffice for describing Hawking radiation
as a robust feature of black holes insensitive to UV physics.
(See Refs.\cite{trans-Planckian-4,Gryb:2018pur} for minority opinions.)
The reasons to dismiss the trans-Planckian problem include the following.
(1) According to the nice-slice argument \cite{Polchinski:1995ta,Giddings},
there is no high-energy event around the horizon,
so the low-energy effective theory remains valid.
(2) The trans-Planckian energy of an outgoing quantum mode is not Lorentz-invariant.
It does not invalidate an effective theory.
(3) Even if we modify the dispersion relation at high energies
(with respect to freely falling frames)
such that there is a UV cutoff in energy,
Hawking radiation remains essentially the same.
\cite{Unruh:1994je,Brout:1995wp,Corley:1996ar,Barcelo:2005fc}.
Other general arguments have also been given for Hawking radiation's independence of UV physics
\cite{Hambli:1995pp,Lubo:2003rs,Unruh:2004zk,Agullo:2009wt,Kajuri:2018myh}.

However,
it was recently found \cite{Ho:2021sbi} that
higher-derivative interactions with the background geometry or collapsing matter
modify Hawking radiation by
an exponentially growing rate of particle production.
(The effect of renormalizable interactions on Hawking radiation
was considered in Ref.\cite{Leahy:1983vb},
and Ref.\cite{Akhmedov:2015xwa} reports
a large one-loop effect on Hawking radiation.)
The origin of this effect is
a Lorentz-invariant trans-Planckian energy
between the outgoing quantum modes and the collapsing matter.
Even though the outgoing modes are still virtual particles at the horizon
(hence the horizon remains uneventful as demanded by the nice-slice argument),
their time evolution through the collapsing matter
relies on trans-Planckian physics \cite{Ho:2021sbi}.
On the other hand,
in agreement with the nice-slice argument,
the horizon is uneventful for freely falling observers
\cite{Ho:2021sbi}.

The goal of this work is to further illuminate
how Hawking radiation relies on UV physics.
We prove in Sec.\ref{HR-spectrum} that
a UV cutoff turns off Hawking radiation after the scrambling time
\footnote{
The phrase ``scrambling time'' was introduced in Ref.\cite{Sekino:2008he}
in a different context.
}
in the free-field approximation.
Since the radiation emitted before the scrambling time is negligible
compared with its total mass,
the black hole is essentially classical.
This is perhaps the simplest resolution
to the black-hole information loss paradox.

In the absence of a UV cutoff,
we show in Sec.\ref{UV-IR} that a higher-derivative interaction
with the collapsing matter (or background geometry)
turns on a novel UV/IR connection.
As a result, 
Hawking radiation is sensitive to IR physics
after a certain critical time.

In Sec.\ref{sec:TimeEvolution},
we reproduce the large amplitude of particle production 
mentioned above \cite{Ho:2021sbi} via a different approach.
We adopt the Hamiltonian formulation
to show how the Unruh vacuum evolves into
a superposition of multi-particle states.
We also point out that, 
normal particles detected by distant observers
have a peculiar feature for freely falling observers
that is crucial for this unexpected large correction to Hawking radiation.

We conclude that Hawking radiation
is not as robust as what many used to believe.
Instead, it is vulnerable to UV and IR physics.


\section{Hawking Radiation With UV Cutoff}
\label{HR-spectrum}


For simplicity,
we consider a Schwarzschild black hole of mass $M$
dimensionally reduced to 2D with the metric
\begin{align}
ds^2 = - \left(1 - \frac{a}{r}\right) dt^2 + \frac{dr^2}{1 - \frac{a}{r}},
\label{Sch}
\end{align}
where $a \equiv 2G_N M$ is the Schwarzschild radius
and $G_N$ the Newton constant.

In this section,
we prove that 
Hawking radiation is turned off after a critical time
by a UV cutoff $\Lam_{\Om} $
on the frequency $\Om$ of outgoing quantum modes
defined with respect to the Kruskal light-cone coordinate $U$.

Albeit not Lorentz-invariant by itself,
the frequency $\Om$ of an outgoing quantum mode defines
a Lorentz-invariant center-of-mass energy $s$
together with the collapsing matter (or the background geometry).
The formation of a black hole demands that
the characteristic momentum for the collapsing matter must be larger than $1/a$.
Thus the center of mass energy satisfies $s \gtrsim \Om/a$.
A cutoff at the Planck scale on $s$ ($M^2_p > s$)
leads to an effective UV cutoff on $\Om$
\be
\Lam_{\Om} \sim (a M_p)^n M_p
\label{LamUV-Mp}
\ee 
for $n = 1$.
The conclusion of our discussions in this work remains valid
even if we have a much higher UV cutoff \eqref{LamUV-Mp} with $n = 2, 3, 4, \cdots$.

After a short review of the exponential blue-shift at the horizon (Sec.\ref{sec:blueshift})
and the notion of wave packets (Sec.\ref{sec:wavepacket}),
we prove in Sec.\ref{Particle-Number} that,
with a UV cutoff $\Lam_{\Om}$ imposed on the radiation field,
Hawking radiation is turned off after the scrambling time
$\sim \mathcal{O}(a \log (a\Lam_{\Om}))$
in terms of the Schwarzschild time coordinate.
We comment on issues related to the UV cutoff
from the perspective of a UV theory in Sec.\ref{Comment-UV}.


\subsection{Blue Shift And Scrambling Time}
\label{sec:blueshift}


In this subsection,
we review a salient feature of black holes ---
the exponentially large blue-shift factor
(or exponentially small red-shift factor, depending on your perspective)
around the horizon.

The retarded light-cone coordinates are
conventionally denoted by $u$ \eqref{def-uv} and $U$ \eqref{U-def}, respectively,
in the Schwarzschild coordinates and the Kruskal coordinates.
(See Appendix \ref{A} for more details.)
In the near-horizon region,
there is an exponential red-shift factor
\be
\frac{dU}{du} \simeq e^{- \frac{u}{2a}},
\label{U-u}
\ee
which is the origin of Hawking radiation.
With a shift in $U$,
we can choose
\be
U \simeq - 2a e^{- \frac{u}{2a}}
\label{U-def-0}
\ee
near the horizon
so that the horizon is located at $U = 0$.
In an order-of-magnitude estimate,
the initial time $u = 0$ is by definition
the moment when the collapsing matter 
is at a distance $\sim \mathcal{O}(a)$
away from the horizon.

As a result of the red shift \eqref{U-u},
in the eikonal approximation,
the frequency $\om$ of an outgoing wave packet 
defined with respect to $u$
corresponds to an exponentially large blue-shifted frequency
\be
\Om \sim \om \left(\frac{dU}{du}\right)^{-1} \simeq \om e^{\frac{u}{2a}}
\label{blue-shift}
\ee
defined with respect to $U$ at late times (large $u$).
For a wave packet with the central frequency $\om \sim \mathcal{O}(1/a)$,
the corresponding blue-shifted frequency $\Om$ is Planckian
(i.e. $\Om \sim \mathcal{O}(M_p)$) when $u \sim 2a \log(a M_p)$.
We refer to it as the {\em scrambling time} \cite{Sekino:2008he}:
\be
u_{scr} \equiv 2a \log(a M_p).
\label{scr-time}
\ee
For every increase in $u$ by a scrambling time,
the blue-shift is a factor of $a M_p$ larger.

In general,
the frequency $\Om$ becomes unimaginably large
within the order of magnitude of the scrambling time
\footnote{
\label{OoM}
We are concerned with large black holes in the limit of large $a M_p$,
and the order of magnitude of a quantity is defined in terms of $a M_p$,
e.g. $\mathcal{O}(a^n M_p^n)$.
By definition,
finite numbers,
$0.01$ and $100$,
are all of $\mathcal{O}(1)$.
}
\be
\Delta u \sim \mathcal{O}(u_{scr}).
\ee
As an example,
for a solar-mass black hole,
$a_{\odot} M_p \sim 10^{38}$,
the scrambling time is $u_{scr} \simeq 0.0017$ sec.
After $10$ times the scrambling time
($u \simeq 0.017$ sec),
the blue-shifted energy \eqref{blue-shift} of
a particle with $\om \sim 1/a_{\odot}$ is of the order of
$10^{342} M_p$!
This example gives us a flavor of the dramatic blue shift
over the time scale of the scrambling time.

Since the blue-shift factor $\left(dU/du\right)^{-1}$ is exponentially larger at late times,
imposing a UV cutoff $\Lam_{\Om}$ on $\Om$
should suppress Hawking radiation at large $u$.
In the remaining of this section,
we shall give a rigorous proof of this claim.


\subsection{Wave Packets}
\label{sec:wavepacket}


To learn how Hawking radiation changes over time,
we need to consider localized wave packets,
instead of universe-filling plane waves.

A localized outgoing wave packet centered around $u = u_0$
can be defined as
\be
\psi_{(\om_0, u_0)}(u) =
\int_{0}^{\infty} \frac{d\om}{\sqrt{4\pi\om}} \,
f_{\om_0}(\om) e^{-i\om (u - u_0)}.
\label{wave-packet}
\ee
via a frequency distribution $f_{\om_0}(\om)$
with a given central value $\om_0$.
The normalization condition for $f_{\om_0}(\om)$ is
\begin{align}
\int_{-\infty}^{\infty} du \, \rho_{(\om_0, u_0)}(u)
= \int_0^{\infty} d\om \, \left|f_{\om_0}(\om)\right|^2
= 1,
\label{normalization}
\end{align}
where
\be
\rho_{(\om_0, u_0)}(u) \equiv
i \left[\psi^{\ast} \left(\frac{\del}{\del u} \psi\right)
- \left(\frac{\del}{\del u} \psi^{\ast}\right) \psi\right]
\label{rho-def}
\ee
is the relativistic density of the wave packet $\psi_{(\om_0, u_0)}$.

For a normalized wave packet with an approximate frequency $\om_0$,
we expect that
\be
\int_{u_0 - \Delta u}^{u_0 + \Delta u} du \, \rho_{(\om_0, u_0)}(u) \simeq 1,
\label{cond-rho}
\ee
for a given width $\Delta u \gg 1/\om_0$ (but still $\Delta u \sim \mathcal{O}(a)$).
\footnote{
For instance, $\om_0 \sim 1/a$ and $\Delta u \sim 100 a$.
See footnote \ref{OoM}.
}

Consider the Hawking radiation of a massless scalar field $\phi$.
In the free-field theory,
$\phi$ can be expanded
in terms of the creation-annihilation operators $(b^{\dag}_{\om}, b_{\om})$ as
\be
\phi(u, v) = \int_0^{\infty} \frac{d\om}{\sqrt{4\pi\om}} \,
\left(b^{\dag}_{\om} e^{i\om u} + b_{\om} e^{-i\om u} + \mbox{ingoing modes}\right)
\label{outgoing-field}
\ee
at large distances.
The creation operator $b^{\dag}_{(\om_0, u_0)}$ 
for a particle with the wave packet $\psi_{(\om_0, u_0)}$ \eqref{wave-packet}
can be defined as
\be
b^{\dag}_{(\om_0, u_0)} \equiv
\int_0^{\infty} d\om \, f_{\om_0}(\om)
e^{i \om u_0} 
b^{\dag}_{\om}.
\label{creation-op}
\ee
One can check that the correspondence between QFT and QM for 1-particle states,
\begin{align}
{}_b\langle 0 | \phi(u) b^{\dag}_{(\om_0, u_0)} | 0 \rangle_b
= \psi_{(\om_0, u_0)}(u),
\label{QFT-QM}
\end{align}
is satisfied.
The creation-annihilation operators
(eq.\eqref{creation-op} and its Hermitian conjugate) satisfy
\be
[b_{(\om_0,u_0)}, b^{\dag}_{(\om_0,u_0)}] = 1
\ee
for a given wave packet \eqref{wave-packet} 
following the canonical commutation relation
$[b_{\om}, b^{\dag}_{\om'}] = \d_{\om\om'}$
and the normalization condition \eqref{normalization}.

Physically, 
particles are detected as localized wave packets $\psi_{(\om_0, u_0)}$,
rather than universe-filling plane waves.
As a measure of the magnitude of Hawking radiation,
we are interested in 
the vacuum expectation value 
\be
\langle 0 | {\cal N}_{(\om_0, u_0)} | 0 \rangle
\label{VEV-N-0}
\ee
of the number operator
\begin{align}
{\cal N}_{(\om_0, u_0)}
&\equiv b^{\dag}_{(\om_0, u_0)}b_{(\om_0, u_0)}
\nn \\
&=
\int_0^{\infty} d\om_1 d\om_2 \,
f_{\om_0}(\om_1) f^{\ast}_{\om_0}(\om_2) \,
e^{i (\om_1 - \om_2) u_0} b^{\dag}_{\om_1} b_{\om_2},
\label{Number}
\end{align}
for the Unruh vacuum $| 0 \rangle$ \eqref{vac}.
We will show in the next subsection that
$\langle 0 | {\cal N}_{(\om_0, u_0)} | 0 \rangle \rightarrow 0$
as $u_0 \rightarrow \infty$ whenever there is a UV cutoff $\Lam_{\Om}$
on the frequency $\Om$.


\subsection{Time-Dependent Particle Number}
\label{Particle-Number}


Our goal in this subsection is to prove that
the magnitude of Hawking radiation \eqref{VEV-N-0}
is suppressed at late times
when a UV cutoff is imposed.

Hawking radiation is a result of the difference in the notion of particles
defined with respect to different choices of time.
Positive-frequency modes ($e^{i\om u}$ with $\om > 0$) defined in terms of $u$
is a linear combination of positive and negative-frequency modes
($e^{i\Om U}$ with both $\Om > 0$ and $\Om < 0$) defined in terms of $U$.
This can be described as a Bogoliubov transformation.
Related formulas needed in our calculation below are given in Appendix \ref{A}.

Using eqs.\eqref{Bogo-1}, \eqref{Bogo-2}, and \eqref{beta} in Appendix \ref{A},
we find
\begin{align}
\langle 0 | {\cal N}_{(\om_0, u_0)} | 0 \rangle
&=
\int_0^{\infty} d\om_1 d\om_2 \,
f_{\om_0}(\om_1) f^{\ast}_{\om_0}(\om_2) \,
e^{i (\om_1 - \om_2) u_0}
\int_{0}^{\Lam_{\Om}} d\Om \,
\beta^{\ast}_{\om_1\Om} \beta_{\om_2\Om}
\nn \\
&=
\int_{-\infty}^{2a\log(2a\Lam_{\Om})} \frac{du}{2a} \,
\left|F_{\om_0}(u - u_0)\right|^2,
\label{VEV-N}
\end{align}
where we have carried out a change of variable $\Om = \frac{1}{2a} e^{u/2a}$ and
\begin{align}
F_{\om_0}(u - u_0) \equiv
\frac{a}{\pi} \int_0^{\infty} d\om
\sqrt{\om} \,
f_{\om_0}(\om) \, 
e^{- \pi a \om}
e^{- i \om (u - u_0)}
\Gamma(i2a\om).
\label{F-def}
\end{align}

To verify the spectrum of Hawking radiation,
we are concerned with the frequencies
\be
\om_0 \sim \mathcal{O}(1/a),
\ee
and the frequency distribution $f_{\om_0}(\om)$ of the wave packet $\psi_{(\om_0, u_0)}$
must be sufficiently narrow,
\be
\Delta\om \ll \om_0.
\ee
We can thus approximate the slowly-varying functions of $\om$
(with a characteristic scale much longer than $\om_0$ or $1/a$)
in the integrand of eq.\eqref{F-def}
by their values at $\om = \om_0$.
Eq.\eqref{F-def} can thus be approximated by
\begin{align}
F_{\om_0}(u - u_0) &\simeq
\frac{2a}{\sqrt{\pi}} \, \om_0 e^{- \pi a \om_0} \Gamma(i2a\om_0)
\psi_{(\om_0, u_0)}(u).
\label{F-2}
\end{align}

Plugging eq.\eqref{F-2} into eq.\eqref{VEV-N}
and using the identity
\be
\left|e^{- \pi a \om_0} \Gamma(i2a\om_0)\right|^2
= \frac{\pi}{a\om_0}
\left(\frac{1}{e^{4\pi a\om_0} - 1}\right),
\label{id-1}
\ee
we find
\begin{align}
\langle 0 | {\cal N}_{(\om_0, u_0)} | 0 \rangle
&\simeq
\frac{1}{e^{4\pi a\om_0} - 1}
\left[
\int_{-\infty}^{2a\log(2a\Lam_{\Om})} du \,
\rho_{(\om_0, u_0)}(u)
\right],
\label{VEV-N-final}
\end{align}
where $\rho$ is the particle number density \eqref{rho-def}.
This explicitly shows the time-dependence of the Hawking radiation due to a UV cutoff
for any given wave packet.

Eq.\eqref{VEV-N-final} is simply the Planck distribution
$1/(e^{4\pi a\om_0} - 1)$
multiplied by the probability of finding the particle
of a given wave packet $\psi_{(\om_0, u_0)}(u)$
in the range
\be
u \in (-\infty, 2a\log(2a\Lam_{\Om})).
\label{range}
\ee
The integral
$\int_{-\infty}^{2a\log(2a\Lam_{\Om})} du \, \rho_{(\om_0, u_0)}(u)$
in eq.\eqref{VEV-N-final} is bounded from above by $1$
and it depends on both the UV cutoff $\Lam_{\Om}$
and the wave packet $\psi_{(\om_0, u_0)}$.
Hawking radiation is suppressed if the UV cutoff is low
or if the wave packet is localized at a large $u_0$.
In order to reproduce the standard Hawking spectrum,
\hide{
\be
\langle 0 | {\cal N}_{(\om_0, u_0)} | 0 \rangle \simeq
\frac{1}{e^{4\pi a\om_0} - 1},
\label{Planck-dist}
\ee
}
we need
\be
\int_{-\infty}^{2a\log(2a\Lam_{\Om})} du \,
\rho_{(\om_0, u_0)}(u) \simeq 1,
\ee
which in turn demands that
\begin{align}
&2a\log(2a\Lam_{\Om}) \gtrsim u_0 + \Delta u,
\end{align}
where $\Delta u$ is the width of the distribution $|\psi_{(\om_0, u_0)}(u)|^2$
(see eq.\eqref{cond-rho}).

On the other hand,
for any finite UV cutoff $\Lam_{\Om}$,
particles detected at sufficiently late times
would have wave packets $\psi_{(\om_0, u_0)}$ with
\begin{align}
&u_0 - \Delta u
\gtrsim 2a\log(2a\Lam_{\Om}),
\label{UV-cond}
\end{align}
so that
\be
\langle 0 | {\cal N}_{(\om_0, u_0)} | 0 \rangle
\propto
\int_{-\infty}^{2a\log(2a\Lam_{\Om})} du \,
\rho_{(\om_0, u_0)}(u) \simeq 0,
\ee
and Hawking radiation disappears.
Since $\Delta u \sim 1/\Delta \om \sim \mathcal{O}(a)$ by assumption,
the condition 
\eqref{UV-cond} tells us that Hawking radiation is turned off
around the scrambling time defined by
\footnote{
We shall refer to both eqs.\eqref{scr-time} and \eqref{scr-Lam}
as the scrambling time since they are of the same order of magnitude
even when $\Lam_{\Om}$ is very different from $M_p$
as in eq.\eqref{LamUV-Mp} for a finite $n$.
}
\be
u_{scr}(\Lam_{\Om}) \equiv 2a\log\left(2a\Lam_{\Om}\right).
\label{scr-Lam}
\ee

The condition \eqref{UV-cond} on the disappearance of Hawking radiation is equivalent to
\be
\frac{1}{2a} e^{\frac{u_0-\Delta u}{2a}}
\gg \Lam_{\Om}.
\label{UV-cond-2}
\ee
It states that Hawking radiation is turned off at the time
when the blue-shifted frequency $\Om$ \eqref{blue-shift}
for the dominant frequency $\om \sim 1/a$ in Hawking radiation
is much larger than the UV cutoff $\Lam_{\Om}$.

In view of the trans-Planckian issue of Hawking radiation,
this conclusion is not surprising.
Yet,
it could be a surprise in view of the results on modified dispersion relations with energy bounds
\cite{Unruh:1994je,Brout:1995wp,Corley:1996ar}.
This is a clear evidence of the fact that 
Hawking radiation is not robust against different UV physics.

Notice that
the cutoff effect would have been completely missed
if we did not consider Hawking radiation in terms of wave packets.
In the literature, 
the Hawking spectrum is often studied in terms of 
$\langle 0 | b^{\dag}_{\om} b_{\om'} | 0 \rangle$,
which would not be suppressed by a cutoff
as long as $\Lam_{\Om} \gg 1/a$.

\begin{figure}
\hskip0em
\centering
\begin{minipage}{.9\textwidth}
  \centering
  \includegraphics[width=.45\linewidth]{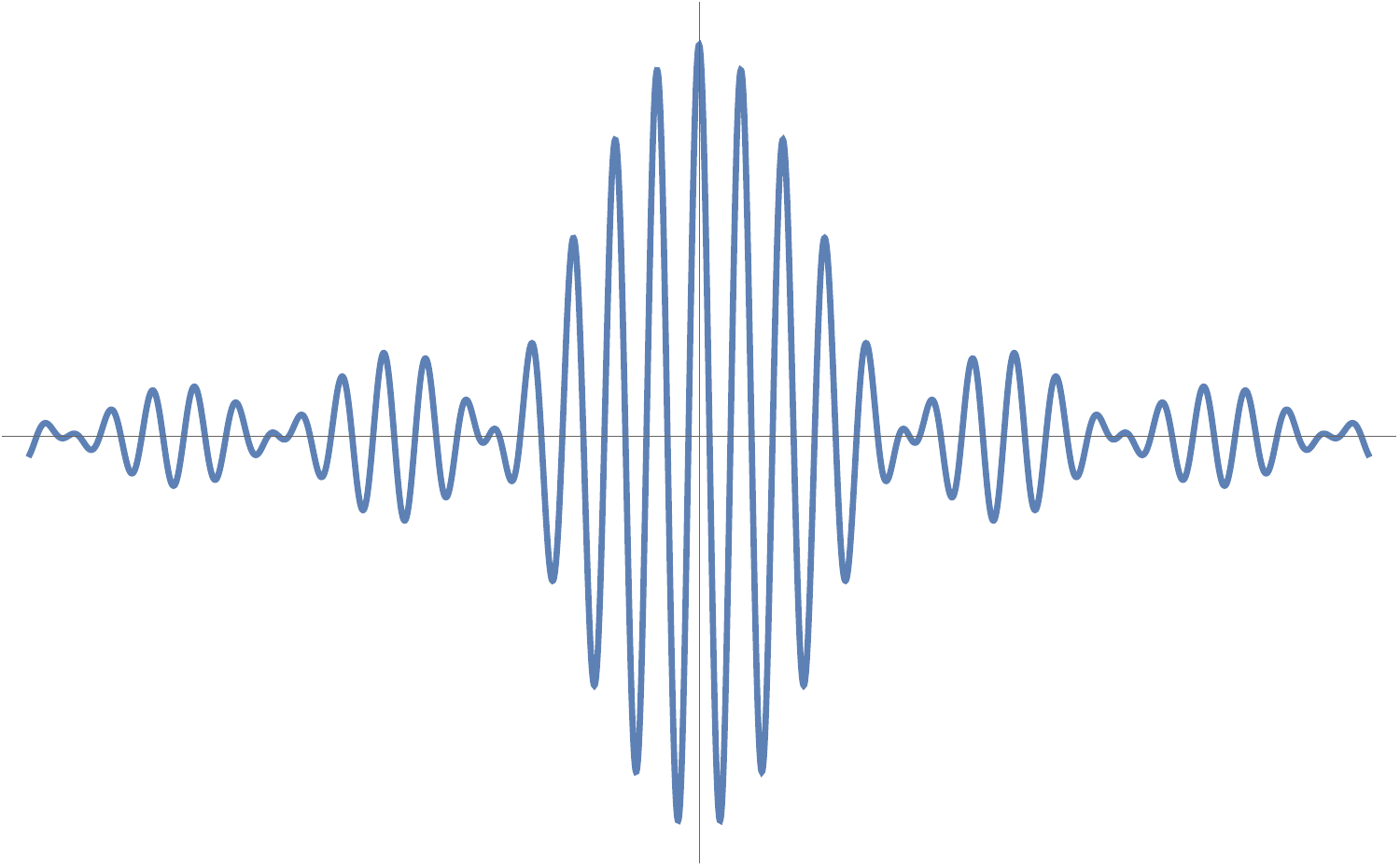}
\vskip0em
\captionof{figure}{\small
The wave packet \eqref{wave-packet}
for the distribution eq.\eqref{f-ex}.
}
\label{fig:wave-packet}
\end{minipage}
\end{figure}

Hawking considered the example \cite{Hawking:1974sw}
\be
f_{\om_0}(\om) 
=
\frac{1}{\sqrt{2\Delta\om}} \left[\Th(\om-\om_0+\Delta\om)
- \Th(\om-\om_0-\Delta\om)\right],
\label{f-ex}
\ee
whose wave packet \eqref{wave-packet} is plotted in Fig.\ref{fig:wave-packet}.
Hawking radiation \eqref{VEV-N-final} is then given by
\begin{align}
\langle 0 | {\cal N}_{(\om_0, u_0)} | 0 \rangle
&\simeq
\left(\frac{1}{e^{4\pi a\om} - 1}\right)
\frac{1}{2\pi\Delta\om[u_0-2a\log(2a\Lam_{\Om})]}
\qquad \mbox{at large $u_0$.}
\label{example}
\end{align}
The condition for this to be small is precisely eq.\eqref{UV-cond}.
Hawking radiation dies off as
\be
\langle 0 | {\cal N}_{(\om_0, u_0)} | 0 \rangle \propto \frac{1}{u_0}
\label{1/u}
\ee
when
\be
u_0 \gg 2a\log(2a\Lam_{\Om}).
\label{latetime}
\ee
The $1/u$ behavior \eqref{1/u} of the Hawking radiation at large $u$ 
is a consequence of the restriction to positive frequencies ($\om > 0$)
in the definition of the wave packet \eqref{wave-packet}.

For $\Lam_{\Om}$ given by eq\eqref{LamUV-Mp},
the mass $\Delta M$ evaporated before the scrambling time is only a negligibly small fraction
\be
\frac{\Delta M}{M} \sim \mathcal{O}\left[\frac{\log(a^2 M_p^2)}{\left(a^2 M_p^2\right)}\right]
\ee
of the total mass $M$.
The energy loss in Hawking radiation can be ignored at the leading order,
and the black hole is well approximated by its classical description.

In Appendix \ref{Schwarzian},
we reach the same conclusion as this subsection in a different way.
Hawking radiation can be derived as an anomalous term
in the energy-momentum tensor of a 2D conformal field theory.
We introduce a UV cutoff in the form of a point-splitting regularization $\d$,
and show that,
when the wave packet is at the separation $\Delta U = \d/2$ from the horizon,
the anomalous term in the coordinate transformation $u \rightarrow U$ vanishes.
We find again that
Hawking radiation is turned off by a UV cutoff at a critical time.


\subsection{Comments on UV Cutoff}
\label{Comment-UV}




In this subsection,
we discuss possible physical realizations of the UV cutoff in a UV theory,
for which laws of physics can be very different from effective theories.
Discussions in this subsection are
more speculative than the rest of the paper.



Although the UV cutoff $\Lam_{\Om}$ is not Lorentz invariant,
it can arise from a Lorentz-invariant cutoff.
For instance,
there can be a UV cutoff on $s$,
the center-of-mass energy between the collapsing matter
and the outgoing quantum modes,
as we mentioned at the beginning of this section.
It is thus possible to have a UV cutoff $\Lam_{\Om}$
in Lorentz-symmetric UV theories.

In general,
the notion of a UV cutoff $\Lam_{\Om}$
corresponds to a minimal length $1/\Lam_{\Om}$ in spacetime.
Interestingly,
a common feature of most (if not all) proposals of quantum gravity
(including string theory) is a minimal length scale
\cite{Amati:1988tn,Scardigli:1999jh,Garay:1994en,Kempf:1994su,Kempf:1996fz,Brau:1999uv,Maggiore:1993rv}
(see Ref. \cite{Hossenfelder:2012jw} for a review),
although it is so far unclear whether it only refers to an effective description,
or to an absence of physical states beyond the cutoff.
Related ideas include noncommutative geometry 
\cite{Snyder:1946qz,Chu:1998qz,Seiberg:1999vs},
the generalized uncertainty principle 
\cite{Kempf:1994su,Scardigli:1999jh},
the doubly special relativity
\cite{Amelino-Camelia:2000stu,Amelino-Camelia:2000cpa}, etc.

If the resolution of the spacetime has a lower bound
\be
\Delta x \gtrsim 1/M_p,
\label{Dx>1}
\ee
beyond which no physical states can be properly defined,
its conjugate momentum $P_x$ must have a UV cutoff.
The projection of the energy $\Om$ of an outgoing mode
on the spatial momentum $P_x$ would be bounded from above.
As long as the projection factor is finite,
a UV cutoff on $\Om$ is implied.




To modify Hawking radiation after the scrambling time,
we only need a condition weaker than eq.\eqref{Dx>1},
for instance,
the spacetime uncertainty principle 
proposed by Yoneya \cite{Yoneya:1987gb,Yoneya:1989ai,Yoneya:2000bt},
which is motivated by string theroy.
This principle is generalized to other spacetime uncertainty relations \cite{Chu:1999wz}
through D-branes and string dualities,
and has been applied to the problem of density fluctuations
in the early universe \cite{Brandenberger:2002nq}.

\hide{
The physics behind the spacetime uncertainty relation \eqref{STUR}
is discussed in Refs.\cite{Yoneya:1987gb,Yoneya:1989ai,Yoneya:2000bt}.
We give a heuristic brief introduction here as follows.
In string theory,
spacetime is not put in by hand.
The definition of spacetime is given in terms of strings
and thus limited by the nature of strings.
When we explore an extremely short period of time $\Delta t \ll \ell_s$,
the usual uncertainty relation in quantum mechanics
demands an extremely large energy $E \gg 1/\ell_s$.
But a string with an extremely large energy has a large spatial extension,
leading to a large uncertainty $\Delta x \sim E/T_s \gg \ell_s$ in space,
where $T_s = 1/(2\pi\ell_s^2)$ is the tension of the string.
Like the uncertainty relation in quantum mechanics,
this is not merely a technical issue that can be avoided
by a clever design of apparatuses.
All properties of the spacetime can only be well defined
to the extent that they can be probed by strings
according this conjecture \eqref{STUR}.
}

Let us now apply
the spacetime uncertainty principle
to Hawking radiation.
In the near-horizon region,
in terms of the local Kruskal coordinate system $(T=(V+U)/2, X=(V-U)/2)$,
we expect the spacetime uncertainty relation 
to be expressed as 
\be
\Delta T \Delta X \gtrsim \ell_s^2
\label{STUR-TX}
\ee
for sufficiently large black holes
in the near-horizon region.
We assume here that $\Delta T$ and $\Delta X$
refer to limits on the spacetime resolution.

First,
the Schwarzschild geometry is meaningless
unless the resolution in space is sufficiently good,
i.e. we must have
\be
\Delta X \lesssim \mathcal{O}(a),
\label{DX}
\ee
otherwise,
we cannot be certain about the formation of a black hole.
The spacetime uncertainty relation \eqref{STUR-TX} then
implies the uncertainty in time
\be
\Delta T \gtrsim \ell_s^2/a.
\ee
Since $U = T - X$,
the uncertainty in $T$ leads to an uncertainty in $U$:
$\Delta U \gtrsim \ell_s^2/a$,
which imposes a UV cutoff
\be
\Lam_{\Om} \lesssim a/\ell_s^2.
\label{UV-cutoff-STUR}
\ee
on the spectrum of $\Om$ in the near-horizon region.
Plugging eq.\eqref{UV-cutoff-STUR} into eq.\eqref{UV-cond}
(and recall that $\Delta u \sim \mathcal{O}(a)$),
we find that Hawking radiation is turned off when
\be
u \gtrsim 4a \log\left(a/\ell_s\right)
= 2u_{scr}(1/2\ell_s),
\label{diamond-cond}
\ee
which is twice the scrambling time $u_{scr}(\Lam_{\Om})$ \eqref{scr-Lam}
for $\Lam_{\Om} = 1/2\ell_s$.
At late times when eq.\eqref{diamond-cond} is satisfied,
the spacetime uncertainty principle prohibits Hawking radiation.

We should comment that the precise interpretation 
of the spacetime uncertainty relation,
and more generally,
that of the minimal length scale in quantum gravity,
is still obscure.
For instance,
a potential interpretation of the minimal length is
not to exclude quantum modes,
but to exclude interactions at a shorter scale.
In this case,
Hawking radiation would not be turned off after the scrambling time.
Regarding such questions,
the information loss paradox could be a guidance
in our search for a self-consistent theory of quantum gravity.


It is of interest to construct explicit toy models with UV cutoffs.
A simple realization of the UV cutoff is to modify
the field operator $\phi$ \eqref{field-a} by setting
\be
[a_{\Om}, a_{\Om'}^{\dag}] = 0
\quad \mbox{for} \quad \Om, \Om' \geq \Lam_{\Om}
\label{aa-cutoff}
\ee
in the infinite past.
The field $\phi$ at a later time is determined by time evolution of the free field.
If we continue to define the operators $(b^{\dag}_{(\om_0, u_0)}, b_{(\om_0, u_0)})$
by eq.\eqref{creation-op} and its Hermitian conjugate
through the mode expansion \eqref{phi-modes} of $\phi$ at large distances,
we have
\begin{align}
[b_{(\om_0, u_0)}, b^{\dag}_{(\om_0, u_0)}] 
&\simeq
\int_{-\infty}^{2a\log(2a\Lam_{\Om})} du \,
\rho_{(\om_0, u_0)}(u).
\label{comm-bb}
\end{align}
In comparison with
the VEV of the particle number \eqref{VEV-N-final},
the UV cutoff not only turns off Hawking radiation,
but it also turns off the commutator \eqref{comm-bb}
even for low-frequency modes
($\om \sim \mathcal{O}(1/a)$)
at the same time.

In App.\ref{B},
we show that a more general modification
of the canonical commutation relations of a field $\phi$
other than the cutoff effect \eqref{aa-cutoff}
leads to the same conclusion that,
in the free-field approximation,
a large modification of the Hawking spectrum implies
a large modification of the canonical commutation relations
for distant observers.


In the next two sections,
we will make the opposite assumptions,
that is,
we will assume no UV cutoff,
but consider the effect of interactions with the background,
and show that Hawking radiation is again sensitive to UV physics.


\section{UV Cutoff From IR Cutoff}
\label{UV-IR}


We show in this section that,
in the absence of a UV cutoff (i.e. $\Lam_{\Om} = \infty$),
due to higher-derivative interactions with matter in a generic effective theory,
outgoing quantum modes at higher frequencies 
are trapped inside the collapsing matter for a longer time.
Unknown physics beyond an IR cutoff (e.g. the age of the universe)
is turned into unknown physics beyond a UV energy scale,
and the effective-theoretic prediction of Hawking radiation becomes unreliable
after a certain critical time.
This is reminiscent of, but different from, the UV/IR connection
in noncommutative field theories \cite{Matusis:2000jf}.


\subsection{Example of Higher-Derivative Interaction}
\label{n=2}


In this section,
we consider the effect of non-renormalizable
--- in particular, higher-derivative --- interactions
on Hawking radiation in an effective theory.
In the context of effective theories,
higher-derivative terms should be treated perturbatively
without enlarging the dimension of the phase space.
(See, for example, Refs.\cite{YF,Cheng:2001du}.)
The effective theory breaks down for its description of a physical effect
when the contribution of non-renormalizable interactions 
dominate over the renormalizable ones.

As an example,
we consider a specific higher-derivative interaction,
\be
{\cal L}_{int} = 
\frac{1}{16}
\frac{\lam_2}{M_p^6} R^{\mu_1\nu_1} R^{\mu_2\nu_2}
(\del_{\mu_1}\del_{\mu_2} \phi) (\del_{\nu_1}\del_{\nu_2} \phi),
\label{Lint}
\ee
between the scalar field $\phi$
and the background Ricci tensor $R_{\mu\nu}$.
It should be clear from the discussion below that
the Ricci tensor is not a crucial ingredient for our consideration.
The same physical effect would arise
if we replace the Ricci tensor by other background fields, 
e.g. $R_{\mu\nu} \rightarrow \del_{\mu}\Phi\del_{\nu}\Phi$,
where $\Phi$ is a background field of the collapsing matter.

For the background under consideration,
the Ricci tensor is dominated by the contribution of the collapsing matter.
For simplicity,
the collapsing matter is assumed to be an ingoing null shell
with the profile
\be
\d_d(V) \equiv \frac{1}{d} \left[
\Th(V + d/2) - \Th(V - d/2)
\right].
\label{dd-def}
\ee
Via Einstein's equations,
the dominant component of the Ricci tensor is
\be
R_{VV} = 8 \pi G_N T_{VV} = \frac{1}{a^3 M_p^2} \, \d_d(V)
\label{RVV-exp}
\ee
up to an overall numerical factor of $\mathcal{O}(1)$.
Using the metric in the near-horizon region \eqref{metric-NHR},
one can show that
the dimensionally reduced interaction term \eqref{Lint} is
\footnote{
In the dimensional reduction from 4D to 2D for spherically symmetric configurations, 
derivatives with respect to the angular variables ($\theta$, $\phi$) can be dropped. 
For our configuration, 
$R_{UU} \simeq 0$ and $R_{UV} \simeq 0$
while $R_{VV}$ is given by eq.\eqref{RVV-exp}.
Using the metric \eqref{metric-NHR},
we can then simplify ${\cal L}_{int}$ \eqref{Lint}.
Finally, 
according to eq.\eqref{dd-def},
we can apply $\delta_{d}(V)\delta_{d}(V ) = \frac{1}{d} \delta_{d}(V)$
to the expression above to derive eq.\eqref{example-L}.
}
\be
{\cal L}_{int} \simeq \frac{\lam_2}{a^6 d M_p^{10}} \, \d_d(V)
(\del_U^2 \phi)^2
\label{example-L}
\ee
at the leading-order of the large-$a$ approximation.

The wave equation for the (dimensionally reduced) action
\be
S = \int dU dV \left(\del_U \phi \del_V \phi + {\cal L}_{int}\right)
\label{example-S}
\ee
is
\be
\del_U\del_V \phi - \frac{\lam_2}{a^6 d M_p^{10}} \, \d_d(V) \del_U^4 \phi = 0.
\label{WE-n=2}
\ee
The general solution to this equation is given by
\begin{align}
\phi = e^{\frac{\lam_2}{a^6 d M_p^{10}} \int_{V_0}^{V} dV' \d_d(V') \del_U^3} \psi(U)
+ F(V)
\label{general-sol}
\end{align} 
for arbitrary functions $\psi(U)$ and $F(V)$.
The first term corresponds to an outgoing wave
and the 2nd an ingoing wave.
The solution of an outgoing wave is only $V$-dependent for $V \in (-d/2, d/2)$.

\begin{figure}
\hskip0em
\centering
\begin{minipage}{.45\textwidth}
  \centering
  \vskip0em
  \includegraphics[width=.9\linewidth]{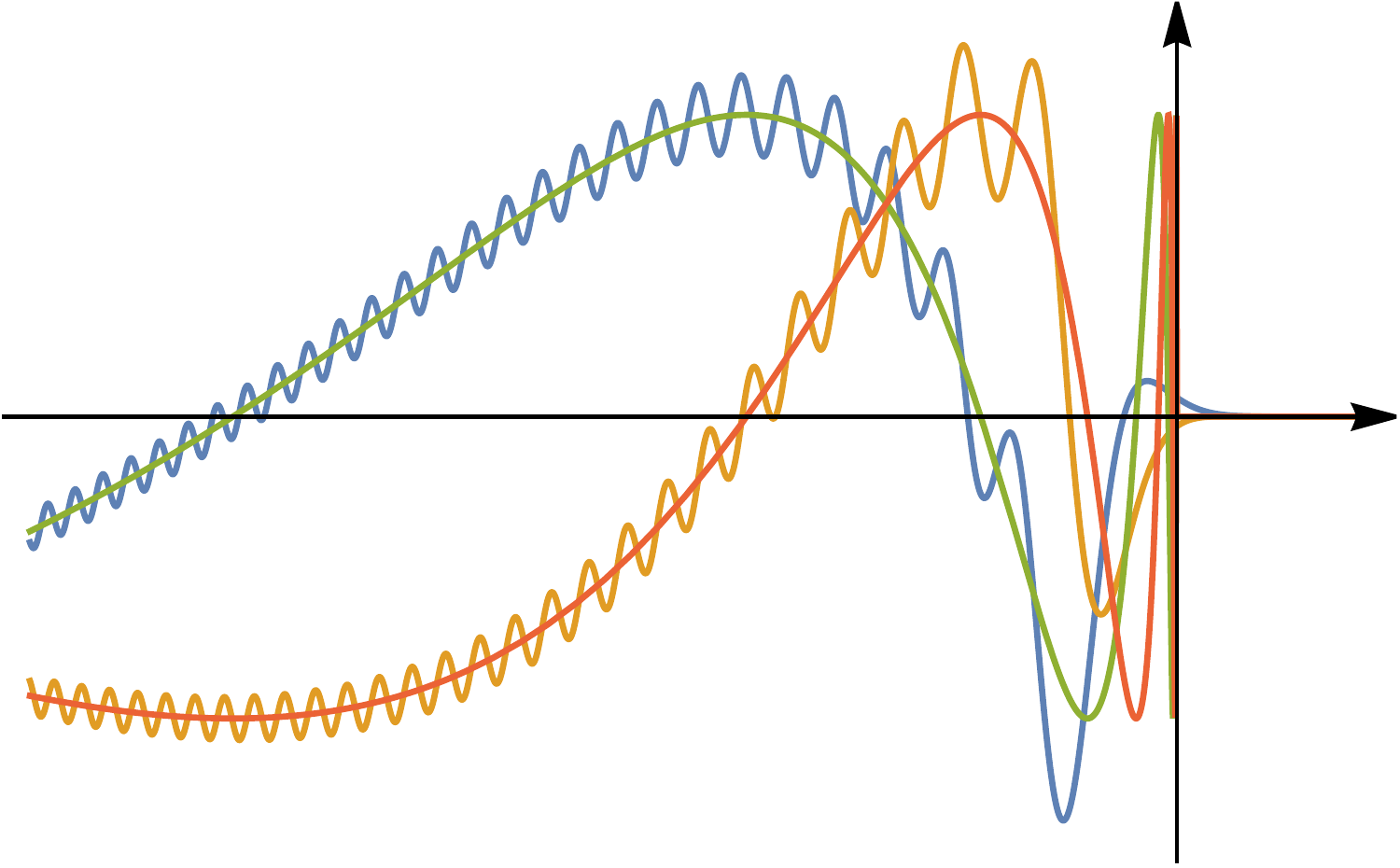}
\vskip.0em
\captionof{figure}{\small
Real and imaginary parts of $\psi(U, V_0)$ and $\psi(U, V_1)$.
$\psi(U, V_0)$ appears as small perturbations of $\psi(U, V_1)$
with ever higher frequencies in the $-U$ direction.
}
  \label{psi-plot-1}
\end{minipage}%
\hskip3.5em
\begin{minipage}{.45\textwidth}
  \centering
  \includegraphics[width=.9\linewidth]{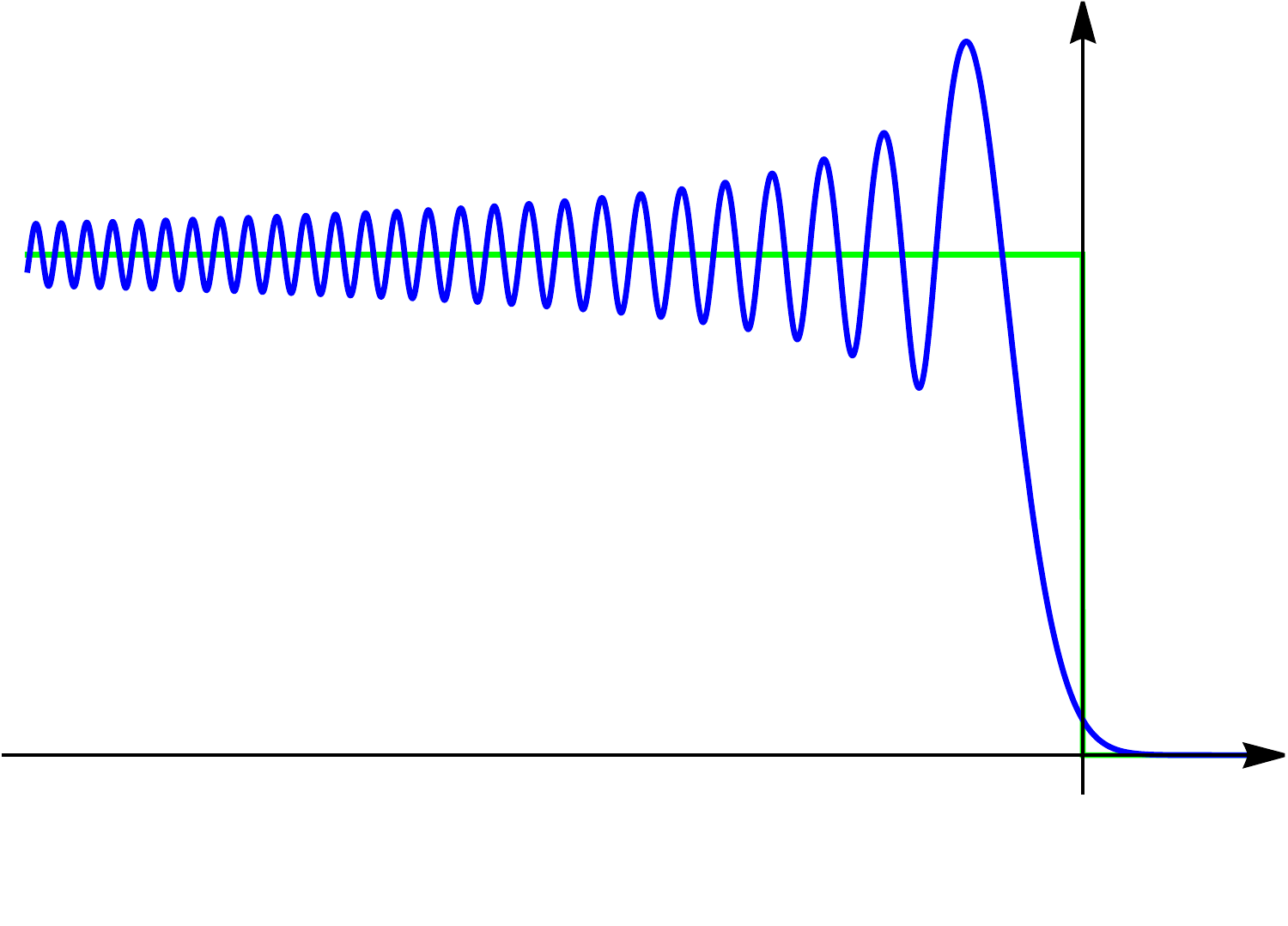}
\vskip-1.5em
\captionof{figure}{\small
Schematic plot of $\left|\psi(U, V_0)\right|$
as a function of $U$ according to eq.\eqref{psi}.
The step function (green) corresponds to the limit $\lam_2 \rightarrow 0$.
}
\label{psi-plot-2}
\end{minipage}
\end{figure}

We use $V_0$ and $V_1$ to denote two points 
inside and outside the collapsing matter, respectively,
so that $V_0 < - d/2$ and $V_1 > d/2$.
In terms of the Kruskal coordinate $U$,
a plane wave for a given frequency $\om$ defined with respect to $u$
for $V > d/2$ has the wave function
\begin{align}
\Psi^{\ast}(U, V_1) 
&= \frac{1}{\sqrt{2\pi\om}} \, e^{i\om u(U)}
= 
c_{\om}
\int_0^{\infty} \, \frac{d\Om}{\Om} \,
(2a\Om)^{i2a\om} e^{i\Om U}
+ \cdots
\label{psi-U}
\end{align}
outside the collapsing matter,
where negative-frequency modes are omitted,
and
\be
c_{\om} \equiv \frac{a}{\pi} \sqrt{\frac{\om}{2\pi}} \, e^{\pi a\om} \Gamma(- i2a\om).
\ee
According to the
general solution to the wave equation \eqref{WE-n=2},
the outgoing mode corresponding to 
eq.\eqref{psi-U} 
is given by
\begin{align}
\Psi^{\ast}(U, V_0)
= c_{\om}
\int_0^{\infty} \, \frac{d\Om}{\Om} \, e^{i\frac{\lam_2}{a^6 d M_p^7}\frac{\Om^3}{M_p^3}}
(2a\Om)^{i2a\om} e^{i\Om U} + \cdots
\label{psi}
\end{align}
inside the collapsing shell.
See Figures \ref{psi-plot-1} and \ref{psi-plot-2} for a plot of $\psi(U, V_0)$.

\begin{figure}
\hskip0em
\centering
\begin{minipage}{.45\textwidth}
  \centering
  \vskip0em
  \includegraphics[width=1.0\linewidth]{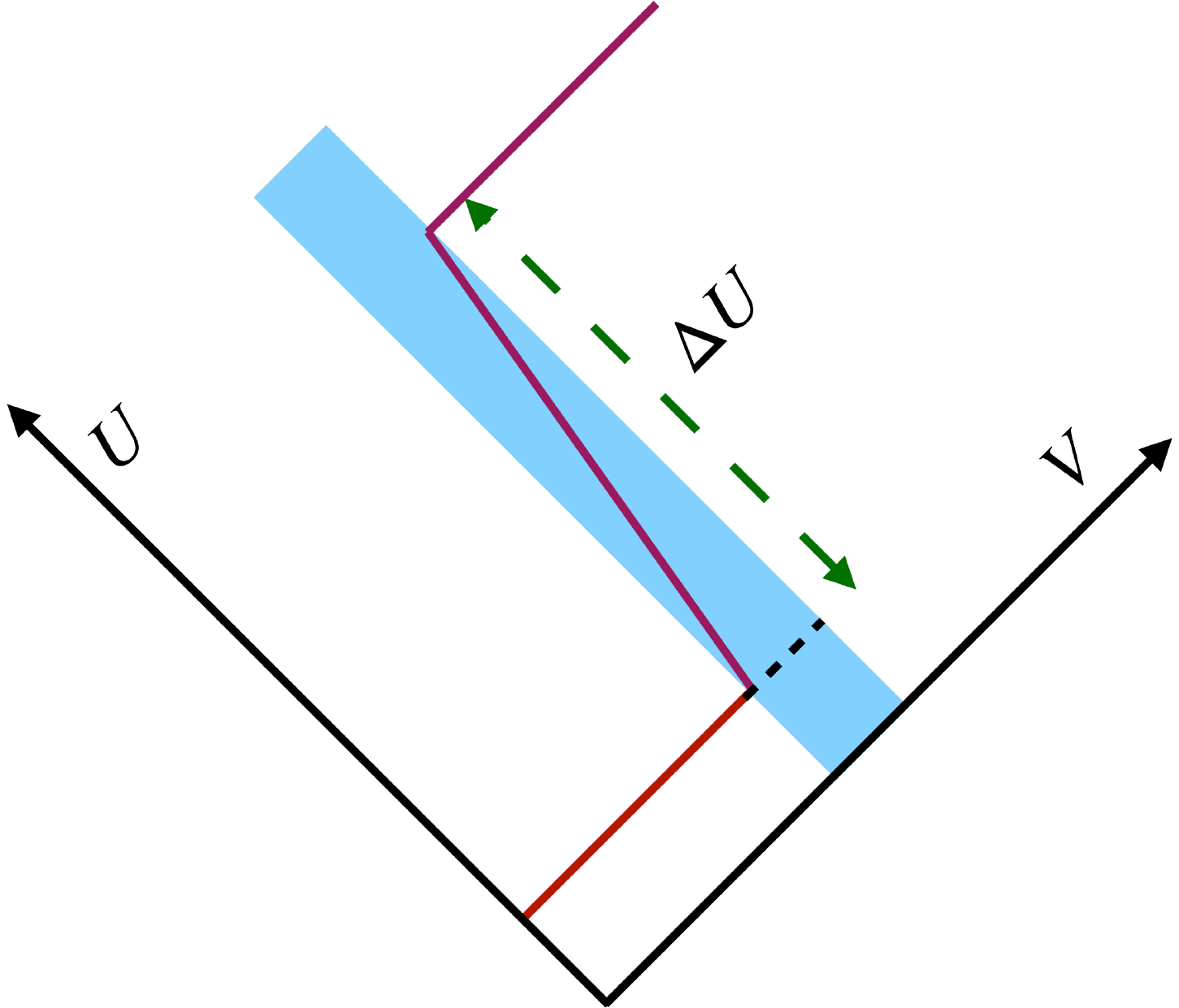}
\captionof{figure}{\small
The trajectory of an outgoing wave packet
with a shift $\Delta U$ \eqref{DeltaU-n=2}
due to its interaction with the collapsing matter (thick blue strip).
}
\label{shift-1}
\end{minipage}
\hskip3.5em
\begin{minipage}{.45\textwidth}
  \centering
  \vskip0em
  \includegraphics[width=.6\linewidth]{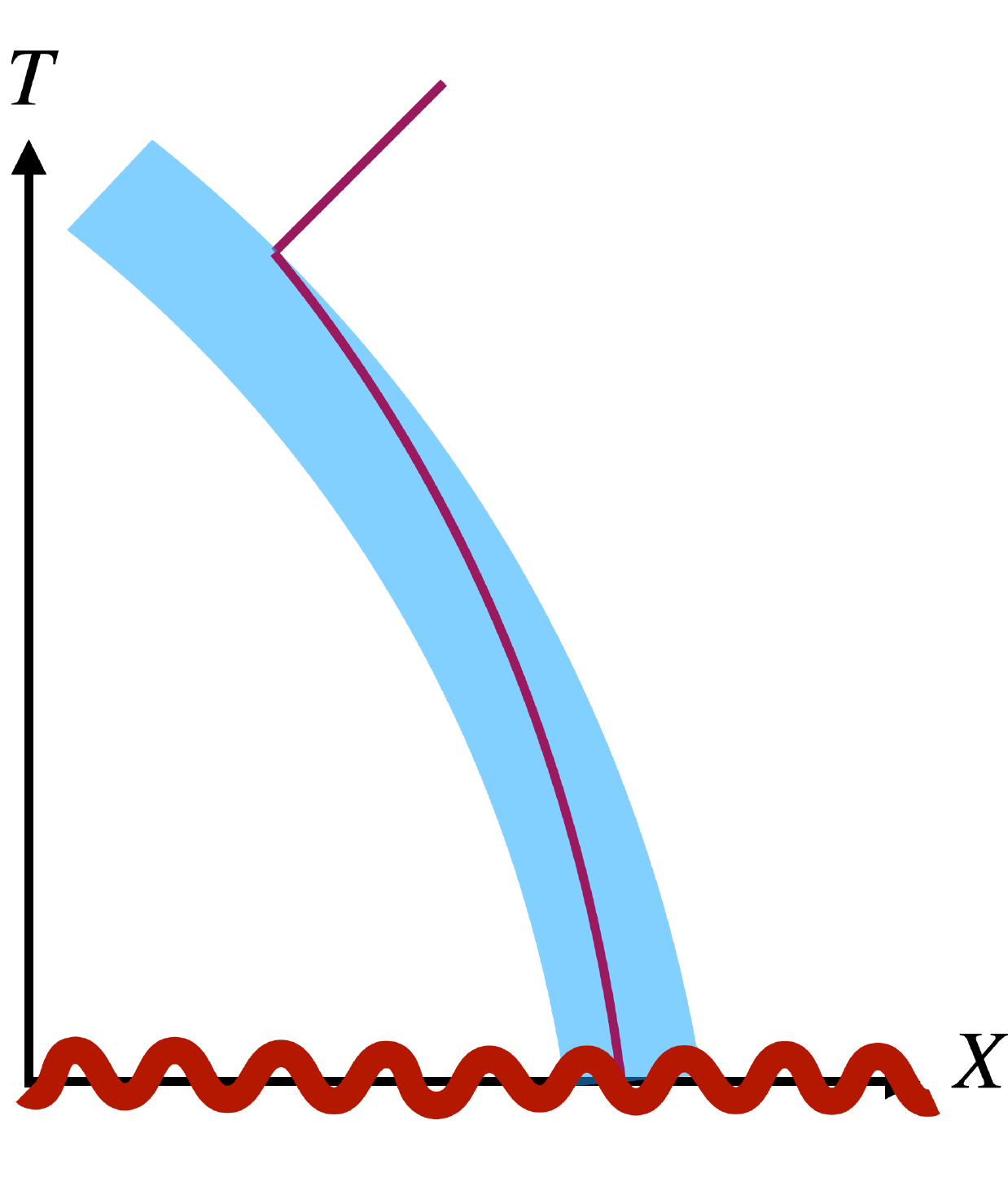}
\vskip0em
\captionof{figure}{\small
The trajectory of a late outgoing wave packet (red curve)
originated from the big bang, 
or any IR cutoff (wavy red curve),
inside a time-like collapsing shell (blue strip).
}
\label{shift-2}
\end{minipage}%
\end{figure}

In the saddle-point approximation,
we take the $\Om$-derivative of the exponent in eq.\eqref{psi}
and find the trajectory of a wave packet to be given by
\be
U 
\simeq - 2a e^{- u/2a} - \Delta U,
\label{U-V0}
\ee
where
\be
\Delta U \equiv \frac{3 \lam_2}{a^6 d M_p^7} \frac{\om^2}{M_p^3} e^{u/a},
\label{DeltaU-n=2}
\ee
using the blue-shift equation \eqref{blue-shift} for $\Om$
in the eikonal approximation.
For an outgoing wave packet centered around a given value of $u$ at large distances,
eq.\eqref{U-V0} gives its $U$-coordinate
when it was inside the collapsing shell.
The first term on the right hand side is simply
the relation \eqref{U-def-0} between the coordinates $U$ and $u$
in the absence of interaction.
The 2nd term is an additional shift due to
the higher-derivative interaction \eqref{example-L} with the collapsing matter.
See Figure \ref{shift-1} for the trajectory of the outgoing wave packet.

For $\om \sim \mathcal{O}(1/a)$ and $d\lesssim a$,
the shift $\Delta U$ dominates over the 1st term in eq.\eqref{U-V0},
when
\be
u \gtrsim \frac{10}{3} a \log\left(a^2 M_p^2\right) = \frac{5}{3} \, u_{scr},
\ee
which is of the same order of magnitude as the scrambling time \eqref{scr-time}.
Over a period of a few scrambling time,
the shift $\Delta U$ can be astronomically large as $\mathcal{O}((aM_p)^n a)$.



\subsection{UV-IR Connection}


As a generalization of the example \eqref{example-L}-\eqref{example-S},
we consider the dimensionally reduced action
\begin{align}
S &\simeq \int dU dV \left[
\del_U\phi \del_V\phi
+ \frac{\lam_k}{(a M_p)^m M_p^{2k-1}} \, \d_d(V) (\del_U^k\phi)^2
\right]
\label{S}
\end{align}
for the Hawking radiation field $\phi$.
The coupling constant $\lam_k$ is dimensionless and by assumption $\mathcal{O}(1)$,
and we have included an additional factor $(a M_p)^m$ for some number $m \geq 0$
to parametrize the strength of the interaction.
The example \eqref{example-L} in Sec.\ref{n=2} corresponds to
the case of $k = 2$ and $m = 7$,
assuming $d \sim \mathcal{O}(a)$.
Despite the fact that the interaction has an extremely small coefficient for $m > 1$,
it has a strong effect on very high-frequency modes
in the absence of a UV cutoff.

Instead of solving wave equations as we did in Sec.\ref{n=2},
here we use the particle picture of wave packets
to reach the same conclusion.
In the saddle-point approximation,
the trajectory of a wave packet is approximated
by that of a particle governed by
the Hamiltonian
\be
H_V = \frac{\lam_k}{(a M_p)^m M_p^{2k-1}} \, \d_d(V) P_U^{2k-1}
\ee
if we treat $V$ as the time coordinate.
\footnote{
By taking the light-cone gauge description,
the issue about instability of the system is omitted in the calculation in this section.
(But it will be touched in the next section.)
We shall focus on the classical effect in this section.
}

The position $U(V)$ and momentum $P_U(V)$ 
along the trajectory of an outgoing wave packet 
obey the Hamilton-Jacobi equations
\begin{align}
\frac{dU(V)}{dV} &=
\{U(V), H_V\} = \frac{(2k-1) \lam_k}{(a M_p)^m M_p^{2k-1}} \, \d_d(V) P_U^{2k-2}(V),
\\
\frac{dP_U(V)}{dV} &=
\{P_U(V), H_V\} = 0.
\end{align}
The trajectory of the particle is subluminal for $(2k-1)\lam_k > 0$.

Since the momentum $P_U$ is a constant along each trajectory,
there is no further mixing of positive and negative frequency modes 
due to the higher-derivative interaction.
One may naively come to the conclusion that 
Hawking radiation is not affected by quadratic interactions
(unless a UV cutoff is imposed by hand).

However,
$dU/dV$ is very large for large $P_U$ within the collapsing matter where $\d_d(V) \neq 0$,
so that the trajectory of the outgoing wave packet is almost light-like in the ingoing direction.
The time evolution of $U(V)$ across the matter shell leads 
to a potentially large shift in $U$:
\begin{align}
\Delta U \equiv U(V_1) - U(V_0)
= \frac{(2k-1) \lam_k}{(a M_p)^m M_p^{2k-1}} P_U^{2k-2},
\label{DeltaU-0}
\end{align}
where $V_0 < -d/2$ and $V_1 > d/2$ are two arbitrary points
inside and outside the collapsing shell, respectively
(see Figure \ref{shift-1}).

Using $P_U \simeq \Om$
for a wave packet of central frequency $\Om$
and eq.\eqref{blue-shift},
we find eq.\eqref{DeltaU-0} equivalent to
\be
\Delta U
\simeq \frac{(2k-1) \lam_k}{(a M_p)^m M_p^{2k-1}} \om^{2k-2} e^{(k-1)u/a}.
\label{DeltaU}
\ee
Eq.\eqref{DeltaU} is in agreement
with the saddle-point approximation \eqref{DeltaU-n=2} in Sec.\ref{n=2}.
Again,
for a given frequency $\om$ at large distances,
$\Delta U$ can be extremely large at late times (large $u$).

For $k > 1$,
the order of magnitude of the shift $\Delta U$ \eqref{DeltaU} is
\be
\Delta U \sim
a \, e^{(k-1)\frac{u - u_{\ast}}{a}},
\label{DeltaT}
\ee
where
\be
u_{\ast} \equiv \frac{2k+m-1}{2(k-1)} \, a \log\left(a^2 M_p^2\right)
\sim \mathcal{O}(u_{scr}).
\label{uast}
\ee

Notice that,
if there is an IR cutoff (e.g. the age of the universe)
on the Kruskal time $T \equiv (U + V)/2$,
it imposes an upper bound $T_0$ on $\Delta U$:
\be
\Delta U \leq T_0.
\ee
Through eq.\eqref{DeltaT},
Hawking radiation is sensitive to the initial condition
at the IR cutoff after $u > u_{\ast} + (a/(k-1))\log(T_0/a)$.
In other words,
it imposes an upper bound on $\Om$ at
\be
\Lam_{\Om} 
\sim \left(a^m M_p^{m+1} T_0\right)^{\frac{1}{2k-2}} M_p,
\label{LamUV}
\ee
in the sense that
an outgoing mode of frequency $\Om \gtrsim \Lam_{\Om}$ 
is originated from within the matter shell at the big bang.
The assumption of the Minkowski vacuum as the initial state
is no longer reliable.

As an example,
for a black hole of the solar mass $M_{\odot}$,
the Schwarzschild radius is $a_{\odot} \sim 3 km$,
and its scrambling time is $\sim 0.0017 \, sec$.
The age of our observable universe is
$T_0 \sim 10^{10} \; \mbox{years}$,
so $\Delta U$ \eqref{DeltaT} becomes larger than $T_0$
when
\be
u - u_{\ast} \gtrsim
a_{\odot} \log\left(\frac{T_0}{a_{\odot}}\right)
\sim 0.0005 \, sec
\ee
for $k = 2$,
where $u_{\ast} = 0.0085 \, sec$
for $m = 7$.
It is smaller for a larger $k$ or a larger Schwarzschild radius $a > a_{\odot}$.
But, roughly speaking,
it is almost always of the same order of magnitude as the scrambling time
$u \sim \mathcal{O}(u_{scr})$.

Conversely,
for $k = 2$,
in order for Hawking radiation to be independent of the IR cutoff
when $u$ reaches $10$ times the scrambling time
$u \simeq 10 a \log\left(a^2_{\odot} M_p^2\right)$,
the IR cutoff
\be
T_0
\gtrsim 10^{1000} \; \mbox{times the size of the observable universe!}
\ee
This bound is even larger for a larger $k$ or a larger $a$.


\subsection{Time-Like Collapsing Matter}


In the above,
we have focused on null collapsing matter for simplicity.
Here we consider the more generic case of collapsing matter
with a time-like trajectory.

For a time-like trajectory of the collapsing matter,
we can go to the local rest frame $(T, X)$ in which the action \eqref{S} becomes
\begin{align}
S &\simeq \int dT dX \left[
(\del_T\phi)^2 - (\del_X\phi)^2
+ \frac{\lam_k}{(a M_p)^m M_p^{2k-1}} \d_d(X) (\del_T^k\phi)^2
\right].
\label{S-timelike}
\end{align}
In terms of the original Kruskal coordinates $(U, V)$,
\begin{align}
T = \frac{1}{2} \left(\xi V + \xi^{-1} U\right),
\qquad
X = \frac{1}{2} \left(\xi V - \xi^{-1} U\right),
\end{align}
where
\begin{align}
\xi &\equiv \sqrt{\frac{dU_s(V)}{dV}}
\end{align}
is given by the velocity of the trajectory $U = U_s(V)$ of the collapsing shell.
We assume that $\xi \sim \mathcal{O}(1)$ for a generic time-like trajectory.

\hide{
The wave equation for the action \eqref{S-timelike} is
\be
(\del_T^2 - \del_X^2)\phi 
- (-1)^k \frac{\lam_k}{(a M_p)^m M_p^{2k-1}} \d_d(X) \del_T^{2k} \phi = 0.
\label{wave-eq-2}
\ee
}
For a sinusoidal wave
$\phi \propto e^{i(ET - PX)}$,
the dispersion relation for the action \eqref{S-timelike} is
\be
E^2 - P^2
+ \frac{\lam_k}{d (a M_p)^m M_p^{2k-1}}\ E^{2k} = 0
\ee
inside the collapsing matter.
The group velocity of an outgoing wave packet
with central momentum $P$ propagating inside the matter is
\begin{align}
W(E) 
= \left(\frac{dP}{dE}\right)^{-1}
= \frac{P(E)}{E + \frac{k \lam_k}{d (a M_p)^m M_p^{2k-1}} E^{2k-1}}.
\label{W-eq}
\end{align}
We assume that $\lam_k > 0$ so that
the wave packet is subluminal ($W(E) < 1$),
and the limit $E \rightarrow \infty$ coincides with $P \rightarrow \infty$.
In the high-energy limit,
eq.\eqref{W-eq} implies that
\begin{align}
W(E) 
\rightarrow 0
\qquad \mbox{as \;\;\;} E \rightarrow \infty
\quad \mbox{for $k > 1$}.
\end{align}

A wave packet of frequency $\om$ at large distances
has a blue-shifted frequency $\Om$ \eqref{blue-shift}.
The continuity of the wave at the surface of the collapsing shell
implies that
\be
E = \xi\Om.
\ee
For a matter shell of thickness $d$,
the time it takes for an outgoing wave packet to pass across the shell is
\begin{align}
\Delta T
&=
\frac{d}{W(E)}
\sim
\frac{a}{(a M_p)^{k+(m-1)/2}} \, e^{(k-1)\frac{u}{2a}}
\end{align}
for $d \sim \mathcal{O}(a)$, $\xi \sim \mathcal{O}(1)$, and $\om \sim \mathcal{O}(1/a)$.

After the scrambling time,
$\Delta T$ becomes extremely large.
As long as the universe has a finite age $T_0$,
all wave packets detected beyond a certain time $u_{cr}$
must be originated within the collapsing shell $(V \in (-d/2, d/2))$
at the initial time of the universe.
(See Figure \ref{shift-2}.)
This critical time is
\be
u_{cr} \equiv \frac{2a}{k-1}
\log\left[
\frac{T_0}{a} \left(a M_p\right)^{k+(m-1)/2}
\right].
\ee

It is assumed in the original derivation of Hawking radiation that
the initial state of the outgoing quantum modes is
the Minkowski vacuum in the infinite past.
This assumption is no longer valid since
these outgoing modes are actually originated from within the matter.
Their initial state is part of the initial state of the collapsing matter.

Unless there is a UV cutoff,
even though the matter can be arbitrarily diluted in the long past,
it always interacts strongly with sufficiently high-frequency modes
through higher-derivative interactions
over the age of the universe.
\footnote{
Even if the matter is much more diluted in the past
so that its Ricci tensor is not $\mathcal{O}(1/a^2)$ but,
say, $\mathcal{O}(1/a^4 M_p^2)$ or $\mathcal{O}(1/a^6 M_p^4)$
in the example of Sec.\ref{n=2},
the critical time would still be of the same order of magnitude as the scrambling time
because of the exponential relation \eqref{U-u}.
}
Therefore,
for Hawking radiation detected after $u = u_{cr}$,
we can no longer assume that
the particles are originated from the Minkowski vacuum,
and the Hawking spectrum is expected to be sensitive to the initial condition.


\section{Particle Creation From Unruh Vacuum}
\label{sec:TimeEvolution}


In the previous section,
we considered quadratic higher-derivative interactions
merely as a modification to the free-field equation.
But it is also possible for them to induce Hawking particle 
\footnote{
By Hawking particles we refer to particles in the spectrum of Hawking radiation.
By definition they have frequencies $\om \sim \mathcal{O}(1/a)$
with a narrow profile $\Delta \om \ll \om$ for distant observers.
}
creation
as they are time-dependent.
In this section, 
we compute the amplitude of particle creation
(for distant observers)
due to higher-derivative interactions with the background
(including the collapsing matter as well as the background geometry).
These non-renormalizable interactions are suppressed by powers of $1/M_p$,
but their contributions to Hawking radiation dominate over the free-field 
around the scrambling time.

Although this amplitude was already calculated in Refs.\cite{Ho:2020cbf,Ho:2020cvn,Ho:2021sbi},
we shall revisit this problem 
by examining the time-evolution of the Unruh vacuum in the Hamiltonian formulation,
so that we can clearly see how the vacuum state evolves towards multi-particle states.
For simplicity, 
we shall adopt the light-cone frame for this calculation.

After a quick review of Hawking radiation in the light-cone frame
in Secs.\ref{FreeField} and \ref{sec:interactions}),
the amplitude for particle creation is computed in Sec.\ref{sec:higher-derivative}.
In Sec.\ref{sec:UV-IR-cutoffs},
we show that Hawking particles
evade the constraint of momentum conservation
due to a peculiar feature of their wave packets,
which makes the large correction to Hawking radiation possible.


\subsection{Hawking Radiation in Light-Cone Frame}
\label{FreeField}


We start with a massless 2D scalar field $\phi$ with the free-field action
\begin{align}
S_0 &\equiv
- \frac{1}{2} \int d^2 x \, \sqrt{-g} \, g^{\mu\nu} \del_{\mu}\phi \del_{\nu}\phi
\simeq \int dU dV \, \del_U \phi \del_V \phi
\label{S0}
\end{align}
in the near-horizon region.
With the Kruskal advanced time coordinate $V$ treated as the time coordinate,
the conjugate momentum of $\phi$ is
\be
\Pi \equiv \del_U \phi,
\label{Pi}
\ee
and the free-field Hamiltonian $H_0$ vanishes.
The momentum in the $U$-direction is
\be
P_U \equiv \int_{-\infty}^{\infty} dU \, \frac{1}{2} (\del_U \phi)^2.
\label{PU}
\ee

In terms of the light-cone coordinate $u$ suitable for distant observers,
only half of the real line of $U$ ($U < 0$) is covered,
and $P_U$ \eqref{PU} becomes
\be
P_U = \int_{-\infty}^{\infty} du \, \frac{1}{2} \, e^{\frac{u}{2a}} (\del_u \phi)^2.
\label{PU-2}
\ee
This is not exactly $P_U$ defined in eq.\eqref{PU},
as it no longer acts on the states behind the horizon ($U > 0$).
Eq.\eqref{PU-2} is more precisely the $U$-momentum operator
after the states behind the horizon are traced over.
Such a trace turns the initial Minkowski vacuum into a density matrix,
and $P_U$ is no longer a conserved quantity.

The periodicity 
\be
u \rightarrow u + i 4\pi a,
\ee
of $P_U$ reflects the fact that
the density matrix is a thermal distribution of $P_U$
at the Hawking temperature
\be
T_H \equiv \frac{1}{4\pi a}.
\ee


\subsection{Turning on Interactions}
\label{sec:interactions}


Now we turn on generic interactions in the effective theory action
\be
S \equiv S_0 + S_{int}^{(0)} + S_{int}^{(1)} + S_{int}^{(2)} + \cdots,
\ee
where
\begin{align}
S_{int}^{(0)} &\equiv
- \int d^2 x \, \sqrt{-g} \, \left(
\frac{1}{2} M^2 \phi^2 + \frac{g}{4!} \phi^4 + \frac{g'}{6!} \phi^6
+ \cdots
\right),
\\
S_{int}^{(1)} &\equiv
\int d^2 x \, \sqrt{-g} \, \left(
\frac{f}{2} \phi^2 g^{\mu\nu} (\nabla_{\mu} \phi) (\nabla_{\nu} \phi)
+ \cdots
\right),
\\
S_{int}^{(2)} &\equiv
\int d^2 x \, \sqrt{-g} \, \left(
\frac{\lam}{8} \, g^{\mu_1\nu_1} g^{\mu_2\nu_2}
(\nabla_{\mu_1}\nabla_{\mu_2}\phi) (\nabla_{\nu_1}\nabla_{\nu_2}\phi)
+ \cdots
\right).
\end{align}
By definition, 
$S_{int}^{(n)}$ includes terms with $2n$ derivatives.
A generic action $S$ should also include interaction terms
between $\phi$ and the collapsing matter,
although they are not explicitly shown in the expressions above.

In terms of the $(u, v)$ coordinate system,
we have
\begin{align}
S_{int}^{(0)} &=
- \int du dv \, \frac{1}{2} \left(1 - \frac{a}{r}\right)
\left(
\frac{1}{2} M^2 \phi^2 + \frac{g}{4!} \phi^4 + \frac{g'}{6!} \phi^6
+ \cdots
\right),
\label{S0-2}
\\
S_{int}^{(1)} &=
- \int du dv \, 
\left(f \phi^2 \del_u \phi \del_v \phi + \cdots\right),
\label{S1-2}
\\
S_{int}^{(2)} &=
\int du dv \, \frac{1}{2} \left(1 - \frac{a}{r}\right)^{-1}
\left(
\lam \left[
(\nabla_u^2\phi) (\nabla_v^2\phi) + (\nabla_u\nabla_v\phi) (\nabla_v\nabla_u\phi)
\right]
+ \cdots
\right).
\label{S2-2}
\end{align}
Due to the measure $\sqrt{-g} = \frac{1}{2}(1-a/r)$,
all terms in $S_{int}^{(0)}$ are negligible in the near-horizon region.
Since Hawking particles detected at large distances at late times
are originated from quantum fluctuations very close to the horizon,
the late-time Hawking spectrum is insensitive to $S_{int}^{(0)}$.
This argument is confirmed in the literature \cite{Leahy:1983vb}.

Terms in $S_{int}^{(1)}$ are not suppressed in the near-horizon region,
but they are suppressed by powers of $1/M_p^2$
as they are non-renormalizable.
Hawking radiation at late times are insensitive to these interaction terms, too.

It is interesting that,
due to the inverse metric $g^{uv} = -2 (1-a/r)^{-1}$
associated with the derivatives $\nabla_u$ and $\nabla_v$,
higher-derivative terms in $S_{int}^{(n)}$
come with the enhancement factor
\be
S_{int}^{(n)} \propto \left(1-\frac{a}{r}\right)^{-(n-1)}
\propto e^{\frac{(n-1) u}{2a}}
\label{enhance}
\ee
in the near-horizon region.
Therefore,
even though they are suppressed by powers of $1/M_p^2$,
all higher-derivative interactions with $n\geq 2$ become important at late times 
(large $u$).

The suppressing powers of $1/M_p^2$ are dominated
by this enhancement factor \eqref{enhance}
around the scrambling time,
and higher-derivative interactions dominate free-field contribution.
The potential importance of higher-derivative terms near the horizon 
was first conjectured in Ref.\cite{Iso:2008sq},
and then confirmed in Ref.\cite{Ho:2020cbf,Ho:2020cvn,Ho:2021sbi}.


\subsection{Particle Creation via Higher-Derivative Interactions}
\label{sec:higher-derivative}


Let us now study the effect of higher-derivative interactions on Hawking radiation
in the Hamiltonian formulation in the light-cone frame.
Since the free-field action $S_0$ has a trivial Hamiltonian $H_0 = 0$,
and $S_{int}^{(0)}+S_{int}^{(1)}$ is negligibly small as we explained above,
the Hamiltonian is dominated by the contribution of $\sum_{k=2}^{\infty} S_{int}^{(k)}$.

We are interested in the energy carried by the outgoing wave packets outside the horizon, 
so our description is restricted to the region outside the horizon
and particles inside the horizon are ignored. 
We emphasize that a calculation of Hawking radiation 
does not need to include the region behind the event horizon. 
(See, e.g., Ref.\cite{HR}.)
In such a calculation of Hawking radiation, 
the energy conservation is not manifest
as the particles behind the horizon are ignored.

As an example,
we consider the higher-derivative coupling to the Ricci tensor given by
\begin{align}
S_{int}^{(k)} &\equiv
\frac{\lam_k}{2^{2k}M_p^{4k-2}}
\int d^2 x \, \sqrt{-g} \, 
g^{\mu_1\nu_1} \cdots g^{\mu_{2k}\nu_{2k}} 
R_{\mu_1\mu_2} \cdots R_{\mu_{2k-1}\mu_{2k}}
\times
\nn \\
&\qquad \times
(\nabla_{\nu_1} \cdots \nabla_{\nu_k}\phi) (\nabla_{\nu_{k+1}} \cdots \nabla_{\nu_{2k}}\phi).
\label{Ricci-coupling}
\end{align}
The coupling constant $\lam_k$ is dimensionless.
Our conclusion would be essentially the same if the Ricci tensor is replaced
by other background fields such as $\del_{\mu}\Phi \del_{\nu}\Phi$,
where $\Phi$ is a matter field in the collapsing matter.

In the near-horizon region where $g_{UV} \simeq -1/2$,
eq.\eqref{Ricci-coupling} is simplified as
\begin{align}
S_{int}^{(k)}
&\simeq
\frac{\lam_k}{2^{2k}M_p^{4k-2}}
\int du dv \, \sqrt{-g}
R_{vv}^k
\left(g^{uv}\right)^{2k}
(\nabla_u^k\phi)^2
\nn \\
&
\simeq
\frac{\lam_k}{2M_p^{4k-2}}
\int dU dV \, 
R_{VV}^k
(\del_U^k\phi)^2,
\label{Sn}
\end{align}
where $R_{\mu\nu}$ is dominated
by the component $R_{VV} \gtrsim \mathcal{O}(1/a^2)$
for a null collapsing matter.
Since eq.\eqref{Sn} does not involve time-derivatives of $\phi$
(i.e. $\del_V\phi$),
the conjugate variable $\Pi$ \eqref{Pi} is not modified
by the interaction \eqref{Ricci-coupling} in the light-cone frame.
The Hamiltonian defined with respect to $V$ is
\begin{align}
H_k(V) &\simeq 
- \frac{\lam_k}{2M_p^{4k-2}} \int
dU
(R_{VV}(U, V))^k (\del_U^k\phi)^2.
\label{HV}
\end{align}

Strictly speaking,
the integrand $(R_{VV}(U, V))^k (\del_U^k\phi)^2$
on the right-hand side of eq.\eqref{HV} should be corrected
by the Schwarzschild geometry outside the near-horizon region,
as the approximation \eqref{Sn} is only valid in the near-horizon region.
However,
such corrections are irrelevant to our calculations below
since we will focus on outgoing wave packets 
originated from the near-horizon region at large $u$.
These wave packets reside well within
a very small range of $U$ close to $U=0$.
The time-evolution of these wave packets is independent of 
the integrand outside this small range in eq.\eqref{HV}.
The detail is given in App.\ref{D}.

This interaction \eqref{HV} induces the transition
from the Unruh vacuum to a 2-partilce state
of certain wave packets $\psi, \psi'$:
\be
| 0 \rangle \quad \longrightarrow \quad
| \psi, \psi' \rangle.
\label{0-2}
\ee
To simplify the derivation, 
we define the final state $|\psi, \psi'\rangle$ as a 2-particle state
(i.e. a superposition of the states $a^{\dag}_{\Om}a^{\dag}_{\Om'}|0\rangle$)
that satisfies the relation
\begin{align}
\langle 0 | \phi(u)\phi(u') | \psi, \psi' \rangle
= \psi(u)\psi'(u) + \psi(u')\psi'(u).
\label{0phiphipsipsi-0}
\end{align}
The explicit construction of this state $|\psi, \psi'\rangle$
is given in App.\ref{D}.
This would greatly simplify the derivation.

In Ref.\cite{Ho:2021sbi},
the final state of the transition \eqref{0-2}
is chosen to be 
\be
b^{\dag}_{(\om_0, u_0)} b^{\dag}_{(\om'_0, u'_0)}|0\rangle,
\ee
where $b^{\dag}_{(\om_0, u_0)}$ is defined by eq.\eqref{creation-op}.
We will see below that the conclusions about
the transition amplitudes for both choices of the final state
are the same:
the amplitude grows exponentially in time
and becomes of order $1$ around the scrambling time.

At the leading order,
the transition amplitude for the process \eqref{0-2} is
\footnote{
Here, we have adopted the formulation of the light-front quantization,
and treated $V$ as the time coordinate.
$H_{k}(V)$ is thus the interacting Hamiltonian for the evolution of $\phi$ in the $V$-direction. 
The wave packets are in general $V$-dependent. 
But since $H_{k}(V)$ is only non-zero inside the collapsing shell, 
an outgoing wave packet becomes $V$-independent in the region $V > d/2$.
This is in agreement with the general solution \eqref{general-sol},
which implies that
$\psi(U, V ) = \psi(U, d/2)$ for all $V > d/2$.
}
\begin{align}
{\cal A}_k(V) \equiv
- i \int_{V_0}^{V} dV' \, \langle \psi, \psi' | H_k(V') | 0 \rangle
\label{Ak-0}
\end{align}
from the initial time $V_0$ to a given time $V > V_0$,
where $V_0$ denotes the initial time inside the collapsing matter shell.

After a straightforward calculation given in App.\ref{D},
for Gaussian wave packets $\psi, \psi'$
centered at two points $u_0$ and $u'_0$,
but of the same central frequency $\om_0$ and width $\Delta u$,
the magnitude of the amplitude ${\cal A}_k(V)$ is given by
\begin{align}
|{\cal A}_k(V)|
&\simeq
\frac{|\lam_k|}{M_p^{4k-2}} \,
\left|\int_{V_0}^{V} dV' \, (R_{VV}(0, V'))^k\right|
\frac{|D^2(k)|}{2 \om_0} \, 
e^{\frac{(2k-1)}{2a}\left(\frac{u_0+u'_0}{2}\right)}
e^{-\frac{(u_0 - u'_0)^2}{4\Delta u^2}}
e^{-\om_0^2\Delta u^2} 
e^{\frac{(2k-1)^2}{(4a)^2}\Delta u^2},
\label{Ak-result}
\end{align}
where
\begin{align}
D(k) \equiv
\left[\Pi_{m=1}^{k}\left(i\om_0 + \frac{m-1}{2a}\right)\right].
\label{D(k)-1}
\end{align}

Let us explain the amplitude \eqref{Ak-result}.
Apart from the coupling constant $\lam_k/M_p^{4k-2}$
and the magnitude of the background $|\int_{V_0}^{V} dV' \, (R_{VV}(0, V'))^k$,
the factor $D^2(k)$ \eqref{D(k)-1} arises from the derivatives $\del_u$.
The factor $1/2\om_0$ arises from the normalization of the two wave packets in the final state.
The exponentially growing factor $e^{\frac{(2k-1)}{2a}(u_0+u'_0)/2}$
is the enhancement anticipated in eq.\eqref{enhance}
(at the midpoint of the two wave packets $u = (u_0 + u'_0)/2$).

Since the interaction is local,
the amplitude is small when the two particles are well separated,
hence we have a large suppression factor $e^{-\frac{(u_0 - u'_0)^2}{4\Delta u^2}}$
at large $|u_0 - u'_0|$.
The factor $e^{-\om_0^2\Delta u^2}$ represents
the suppression for the violation of momentum conservation
by the amount $2\om_0$ in the transition.
\hide{
(The last factor $e^{\frac{(2k-1)^2}{(4a)^2}\Delta u^2}$
is an additional enhancement due to higher-derivatives
that we did not expect.)
}

For an order-of-magnitude estimate of ${\cal A}_k$,
we recall that $\lam_k \sim \mathcal{O}(1)$ (by assumption), 
$\om_0 \sim \mathcal{O}(1/a)$ for Hawking particles,
$\Delta u \gg a$ for a narrow profile ($\Delta \om \ll \om_0$),
and that,
for the background under consideration,
\be
\int_{V_0}^{V} dV R^k_{VV} \gtrsim \mathcal{O}(1/a^{2k-1})
\quad \mbox{for large $V$}.
\label{intRVV}
\ee
Our estimate of the amplitude \eqref{Ak-result} at large $V$ is thus
\begin{align}
\left| {\cal A}_k \right| 
\hide{
&\simeq
\frac{|\lam_k|}{M_p^{4k-2}} 
\left|\int_{V_0}^{V} dV' \, R_{VV}(0, V')^k\right|
\frac{|D^2(k)|}{2 \om_0} \, 
e^{\frac{(2k-1)}{2a}u_0}
e^{- \om_0^2\Delta u^2} 
e^{\frac{(2k-1)^2}{(4a)^2}\Delta u^2} 
\nn \\
}
&\gtrsim
\mathcal{O}\left(\frac{1}{(a M_p)^{4k-2}} \, 
e^{\frac{(2k-1)}{2a}u_0}
\right)
\label{A-k-0}
\end{align}
for $\om_0 \leq (2k-1)/(4a)$.
This amplitude ${\cal A}_k$ is greater than $\mathcal{O}(1)$ when
\be
u_0 \gtrsim 2a \log\left(a^2 M_p^2\right)
\sim \mathcal{O}(u_{scr}).
\label{u0>usr}
\ee

In the absence of a UV cutoff,
the derivative expansion of the amplitude ${\cal A}(V)$
may or may not diverge depending on the coupling constants $\lam_k$.
It either predicts the breakdown of the theory
or a (potentially large) correction to Hawking radiation.

So far we have only considered quadratic interactions.
It is straightforward to see that,
in general,
an $n$-point higher-derivative interaction leads to
an exponentially large amplitude
for the transition from the Unruh vacuum to an $n$-particle state
at the leading order.

In the above,
we only considered the contribution of
the collapsing matter to the Ricci tensor.
But in fact, 
the Ricci tensor due to the vacuum stress tensor
is already sufficient for a large correction to Hawking radiation.
As energy conservation demands a negative ingoing energy flux
to match the energy loss in Hawking radiation,
the vacuum has $R_{VV} \sim \mathcal{O}(1/a^4 M_p^2)$
outside the black hole \cite{Davies:1976ei}.
It is much smaller than the contribution of the collapsing matter,
but it also leads to a large correction to Hawking radiation
after $u_0 \gtrsim \mathcal{O}(u_{scr})$.


\subsection{Wave Packets vs Momentum Conservation}
\label{sec:UV-IR-cutoffs}


The reader may find the result in the previous subsection surprising.
The interaction \eqref{Sn} has approximate local translation symmetry in $U$
at the leading order of the $1/a$ expansion.
How is it possible for a vacuum state with $P_U = 0$
to evolve into a state with large $P_U$ with a large probability?

It was explained in Ref.\cite{Ho:2021sbi} that
the momentum conservation of $P_U$ is only approximately observed
in a way compatible with the uncertainty relation
$\Delta P_u \Delta P_U \gtrsim P_u e^{u/2a}/a$,
since $P_u$ and $P_U$ do not commute.
The uncertainty relation admits
a large violation of momentum conservation of $P_U$ at large $u$.

A new insight to this unexpected result is that
it is associated with a peculiar feature of
Hawking particles (or similar wave packets).

To be concrete,
consider an $n$-point higher-derivative interaction 
\be
{\cal H}_{mn}(V) = \frac{h_{mn}(V)}{n!} (\del_U^m \phi)^n
\ee
in the Hamiltonian in the light-cone frame,
where $h_{mn}(V)$ represents the coupling to the background.
(We have focused on $n=2$ terms in the above;
see eq.\eqref{HV}.)
For given wave packets $\Psi_i(U)$ ($i = 1, 2, \cdots, n$),
we define corresponding creation operator $a_i$ such that
\be
\langle 0 | \phi(U) a^{\dag}_i | 0 \rangle = \Psi_i(U).
\ee
Its leading-order contribution to the amplitude of a transition
(as a generalization of eq.\eqref{0-2})
\be
| 0 \rangle \quad \rightarrow \quad
\Pi_{i=1}^n a^{\dag}_{i} | 0 \rangle
\ee
from the vacuum to an $n$-particle state is
\begin{align}
{\cal A}_n &=
\lim_{\Lam, L \rightarrow \infty}
\int_{V_0}^{V_1} dV
\int_{-L}^{L} dU \,
\langle 0 | a_{1} \cdots a_{n} {\cal H}_{mn}(V) | 0 \rangle
\nn \\
&\propto
\lim_{\Lam, L \rightarrow \infty}
\int_{-L}^{L} dU \,
\Pi_{i=1}^{n} \del_U^m\Psi^{\ast}_i(U).
\label{An-1}
\end{align}
Strictly speaking,
the singularity at the origin $r = 0$ of the 3D space
imposes an upper bound on $U$ at $U = 2a$,
so the integral $\lim_{L\rightarrow\infty}\int_{-L}^{L} dU$ in eq.\eqref{An-1}
can be replaced by $\lim_{L\rightarrow\infty}\int_{-L}^{2a} dU$.
But this replacement is irrelevant as the wave packets $\Psi(U)$
of Hawking particles are restricted to $U < 0$.
We can thus keep the integral $\lim_{L\rightarrow\infty}\int_{-L}^{L} dU$ in eq.\eqref{An-1} 
for Hawking particles.

We can then use the identity of the $\d$-function
\be
\lim_{L \rightarrow \infty} \frac{2\sin(\Om L)}{\Om}
= 2\pi \d\left(\Om\right)
\label{sinc-delta}
\ee
so that,
if we take the limit $L \rightarrow \infty$ first,
the amplitude \eqref{An-1} becomes
\begin{align}
{\cal A}_n 
&\propto
\lim_{L \rightarrow \infty}
\int_0^{\infty} \Pi_{i=1}^{n} d\Om_i \,
\Om_i^{m} \tilde{\Psi}^{\ast}_i(\Om_i) \,
\frac{\sin(\sum_i\Om_i L)}{\sum_i\Om_i}
\label{An-2}
\\
&\propto
\int_0^{\infty} \Pi_{i=1}^{n} d\Om_i \,
\Om_i^{m} \tilde{\Psi}^{\ast}_i(\Om_i) \,
\d\left(\sum_i \Om_i\right),
\label{Limit-IR-UV}
\end{align}
where $\tilde{\Psi}_i$ denotes the Fourier transform of $\Psi_i$.
Since $\Om_i \geq 0$ for all $i$,
the $\d$-function $\d\left(\sum_i \Om_i\right)$ vanishes
unless $\tilde{\Psi}_i(\Om_i)$ includes a $\d$-function at $\Om_i = 0$.
As a result,
the momentum conservation is imposed via the $\d$-function in eq.\eqref{Limit-IR-UV},
and the amplitude ${\cal A}_n$ of $n$-particle creation vanishes.
This is the mathematics behind our intuition about momentum conservation.

However,
eq.\eqref{Limit-IR-UV} is clearly incompatible with the large amplitude of particle creation
in Sec.\ref{sec:higher-derivative} that violates $P_U$-conservation.
The origin of the conflict is that the identity \eqref{sinc-delta} 
is only valid as a distribution, not as a function.
If we multiply both sides of eq.\eqref{sinc-delta} by a smooth function
and integrate over $\Om$,
the identity holds.
But the wave packets of Hawking particles 
are not smooth enough for the identity \eqref{sinc-delta} to hold.

Before taking the limit $L \rightarrow \infty$,
eq.\eqref{sinc-delta} is a good approximation
when it is multiplied by a sufficiently smooth function $g(\Om)$
in an integral over $\Om$.
The smoothness condition on  $g(\Om)$ includes,
for instance,
\begin{align}
L \gg \left|g^{-1}(\Om)\frac{dg(\Om)}{d\Om}\right|.
\label{L>dg}
\end{align}
The function $g(\Om)$ relevant to eq.\eqref{An-2} is
\be
g(\Om) = 
\int_0^{\infty} \Pi_{i=1}^{n} d\Om_i \,
\Om_i^{m} \tilde{\Psi}^{\ast}_i(\Om_i) \,
\d\left(\Om - \sum_i \Om_i\right),
\ee
where the wave packets of Hawking particles is
a superposition of the single-frequency modes \eqref{psi-U},
\begin{align}
\tilde{\Psi}^{\ast}_{\om_i}(\Om_i) \propto
\Om_i^{-1+i2a\om_i} \, \Th(\Om_i).
\label{tildepsii}
\end{align}
Note that this function is ill-defined at $\Om_i = 0$.
Correspondingly,
\begin{align}
\frac{dg(\Om)}{d\Om} 
\propto \Om^{nm - 2 + i2a\sum_i \om_i}
\end{align}
and so
\begin{align}
g^{-1}(\Om)\frac{dg(\Om)}{d\Om} \propto \frac{1}{\Om},
\label{1overOm}
\end{align}
implying that the condition \eqref{L>dg} is always violated at $\Om = 0$.
Therefore,
we can not use eq.\eqref{sinc-delta} as a good approximation
for any finite $L$.

Note that this singularity \eqref{1overOm} at $\Om = 0$
appears even when $m = 0$
(without higher-derivatives).
In fact it is closely related to the origin of Hawking radiation.
Recall that the Hawking temperature can be derived from the ratio
of the Bogoliubov coefficients $\a_{\om\Om}$ and $\b_{\om\Om}$,
and this ratio is determined by an analytic continuation of
the factor $\Om^{i2a\om}$ from $\Om > 0$ to $\Om < 0$:
\be
\left|\frac{\b_{\om\Om}}{\a_{\om\Om}}\right|^2
= \left|\frac{(-|\Om|)^{i2a\om}}{(|\Om|)^{i2a\om}}\right|^2 = e^{- 4\pi a\om}.
\ee
Reading this factor as the Boltzman factor $e^{- \om/T_H}$ 
gives the Hawking temperature $T_H = 1/4\pi a$.
What we see here is that
the singularity at $\Om = 0$ of the Hawking particles
that leads to Hawking radiation
also ensures that the identity \eqref{sinc-delta} does not apply.
As a result,
momentum conservation is not imposed on the amplitude,
and it is possible to have a large value for ${\cal A}_n$
as we have seen in Sec.\ref{sec:higher-derivative}.

\hide{
Since all $\Psi_i(U)$'s are non-zero only in the domain $U \in (-\infty, 0]$,
we can replace the integral $\int_{-L}^{L} dU$ in eq.\eqref{An-1}
by $\int_{-L}^0 dU$ and take the limit $L \rightarrow \infty$.
For $\tilde{\Psi}_i$ given by $\tilde{\Psi}_{\om_i}$ \eqref{tildepsii},
the amplitude ${\cal A}_n$ can directly evaluated as
\begin{align}
{\cal A}_n &\propto
\lim_{\Lam, L \rightarrow \infty} 
\int_0^{\Lam} \Pi_{i=1}^{n} d\Om_i
\frac{\Om_i^{m-1+i2a\om_i}}{\sum_j \Om_j}
\left(1 - e^{- i \sum_{i=1}^{n} \Om_i L}\right)
\nn \\
&=
\lim_{\Lam \rightarrow \infty} 
\Lam^{nm - 1 + i2a\sum_j \om_j}
\left[
\int_0^1 \Pi_{i=1}^{n} dx_i \,
\frac{x_i^{m-1+i2a \om_i}}{\sum_j x_j}
\right],
\label{An-div}
\end{align}
which diverges whenever $nm > 1$.
\footnote{
The quantity in the square bracket is in general finite.
But it is ill-defined for $n = m = 1$.
}
The divergence disappears for suitable wave packets $\Psi_i$
as superpositions of single frequency modes \eqref{tildepsii}.
We can then take the limit $\Lam \rightarrow \infty$
and evaluate ${\cal A}_n$.
Since momentum conservation is no longer imposed,
it is possible to have a large value for ${\cal A}_n$
as we have seen in Sec.\ref{sec:higher-derivative}.
}


\section{Conclusion}
\label{conclusion}


\hide{
[Dilemma:
If there is a UV cutoff,
QFT seems incomplete (?)
and it is hard to imagine how the cutoff is imposed
in a generally covariant way.
But if there is no cutoff,
there are always extremely high frequency modes
which turns any tiny background into a large effect
when this is a higher-derivative interaction.
If the higher-derivative interactions turns off 
strong effect of the background,
does it not also remove the high-frequency modes
in a generally covariant theory?]
}

Contrary to the common belief,
we found that Hawking radiation is sensitive to certain UV as well as IR physics.
Regardless of whether there is a UV cutoff,
the effective theory breaks down around the scrambling time
for its prediction on Hawking radiation;
a UV theory for trans-Planckian physics is needed.

Concrete examples of UV and IR effects on Hawking radiation
are demonstrated in this paper.
First,
a UV cutoff turns off Hawking radiation around the scrambling time.
This is perhaps the simplest resolution to the information loss paradox.

On the other hand,
in the absence of a UV cutoff,
higher-derivative interactions
have large effects on Hawking radiation after the scrambling time.
For instance,
outgoing quantum modes at very high energies 
can be trapped inside the matter for the age of the universe;
it is no longer justified to assume the Minkowski vacuum
to be their initial states.

Furthermore,
higher-derivative interactions induce
a large correction to Hawking radiation
through an exponentially growing particle production for distant observers.
The approximate momentum conservation of $P_U$
expected in the near-horizon region of a large black hole
does not impose a stringent constraint on Hawking particles
due to a peculiar feature of their wave packets.

Since the prediction of an effective theory is valid only when
the contributions of the ignored terms are small,
the large effect of higher-derivative (non-renormalizable) interactions imply that
the effective-theory predictions about Hawking radiation
is unreliable after the scrambling time.

Despite the dependence of Hawking radiation on unknown UV and IR physics,
our results are consistent with the nice-slice argument \cite{Polchinski:1995ta,Giddings}.
For freely falling observers,
the horizon remains uneventful as long as 
they are concerned with ``normal'' wave packets in their local frame.
Particles created with a large production rate are ``normal'' to distant observers
but they have exotic wave packets \eqref{psi-U} for freely falling observers,
as the coordinate transformation \eqref{U-def-0}
maps a normal wave packet in $u$ to an exotic one in $U$.

The reason why particles with wave packets of different nature can have very different physics
(e.g. different production rates)
is that physical observations are always limited by uncertainty relations,
which impose different constraints on different wave packets.
For instance,
if an observation is carried out within an extremely short time scale $\Delta t$,
it is possible to detect particles even in the Minkowski vacuum
that violates energy conservation by an amount $\Delta E \sim 1/\Delta t$.
It was shown in Ref.\cite{Ho:2021sbi}
that the large corrections to Hawking radiation described in Sec.\ref{sec:higher-derivative}
is compatible with the uncertainty relation.

Although Hawking particles are not recognized as ``normal particles''
for freely falling observers so that
they do not observe a large particle production rate \cite{Ho:2021sbi}
(unless they conduct experiments at a trans-Planckian time scale),
it remains to be seen whether the higher-derivative interactions still contribute
a large energy-momentum tensor around the horizon.
This could significantly modify the background geometry
and invalidate the analysis in this paper,
where the back reaction is assumed to be small.
The possibility that a large energy-momentum tensor around the horizon
has been studied in the literature by taking into consideration
of the back-reaction of the quantum matter fields
through semi-classical gravity \cite{Kawai:2013mda,Kawai:2014afa,Kawai:2020rmt,Kawai:2021qdk}.

As we have refuted the robustness of Hawking radiation in effective theories,
the information loss paradox is no longer a paradox of effective theories.
It would remain uncertain how information is preserved
until we analyze the problem in a UV theory.
It will be of interest to explore UV theories,
even toy models,
how different types of UV or IR physics modify Hawking radiation.
The information loss paradox may serve as a guideline
in our pursuit for the theory of quantum gravity.


\section*{Acknowledgement}

We thank Ronny Chau, Hsien-chung Kao, Suguru Okumura, Wei-Hsiang Shao, and Cheng-Tsung Wang
for valuable discussions. 
P.M.H. is supported in part by the Ministry of Science and Technology, R.O.C.
(MOST 110-2112-M-002-016-MY3).
H.K. thanks Prof. Shin-Nan Yang and his family
for their kind support through the Chin-Yu chair professorship,
and is partially supported by Japan Society of Promotion of Science (JSPS),
Grants-in-Aid for Scientific Research (KAKENHI) Grants No.\ 20K03970 and 18H03708,
and by the Ministry of Science and Technology, R.O.C. (MOST 111-2811-M-002-016).


\appendix


\section{Review of Hawking Radiation}
\label{A}

For a more comprehensive review of Hawking radiation,
see Ref.\cite{Brout:1995rd}.
Including the spherical part $ds^2_{S^2} = r^2 (d\th^2 + \sin^2\th d\varphi^2)$,
the Schwarzschild metric \eqref{Sch} can be rewritten as
\begin{align}
ds^2 
&= - \left(1 - \frac{a}{r}\right) du dv + r^2(u, v) (d\th^2 + \sin^2\th d\varphi^2),
\label{Schwarzschild-metric}
\end{align}
where the retarded and advanced Eddington-Finkelstein light-cone coordinates are
\begin{align}
u \equiv t - r_{\ast}, 
\qquad
v \equiv t + r_{\ast},
\label{def-uv}
\end{align}
and the tortoise coordinate is defined by
\begin{align}
r_{\ast} \equiv r - a + a \log\left(\frac{r}{a} - 1\right).
\end{align}

The Schwarzschild metric is modified by
the collapsing matter as well as 
the vacuum expectation value of the energy-momentum tensor
(including Hawking radiation).
Assuming certain upper bounds on the vacuum energy-momentum tensor
such that the horizon is uneventful,
the modification to the metric through 
the semi-classical Einstein equation
is analyzed in Refs.\cite{Ho:2019pjr,Ho:2019qiu},
which also explain how to use the Schwarzschild metric as an approximation.
Our use of the Schwarzschild metric in this paper
is compatible with these works \cite{Ho:2019pjr,Ho:2019qiu}.

In the near-horizon region,
the spacetime is nearly flat for a large black hole.
We can approximate the metric by
\be
ds^2 \simeq - dU dV + r^2 (d\th^2 + \sin^2\th d\varphi^2),
\label{metric-NHR}
\ee
where
\begin{align}
U(u) &\simeq - 2a e^{- \frac{u}{2a}},
\label{U-def}
\\
V(v) &\simeq 2a e^{\frac{v}{2a}}.
\label{V-def}
\end{align}
The future horizon at $u = \infty$ corresponds to $U = 0$.
The light-cone coordinates $(u, v)$ are convenient for distant observers,
and the Kruskal light-cone coordinates $(U, V)$ 
for freely falling observers around the horizon.

Across the collapsing shell,
the metric \eqref{metric-NHR} continues to the Minkowski metric
\begin{align}
ds^2 &= - dUdV + R^2(U, V) (d\th^2 + \sin^2\th d\varphi^2),
\label{Mink}
\end{align}
inside the collapsing shell,
where the radius is
\begin{align}
R(U, V) \equiv \frac{V - U}{2}.
\label{R}
\end{align}
We patch the metrics \eqref{metric-NHR} and \eqref{Mink}
along the trajectory of the collapsing null shell
at $v_s = 0$,
or equivalently,
$V_s = 2a$.

We can carry out a shift of the coordinate $V$ such that
the collapsing matter is located at a different value, say, $V_s = 0$,
as we did in the main text of this paper.

The $s$-wave modes of a 4D massless scalar field $\phi$ can be expressed as
\begin{align}
\phi &\simeq \int_0^{\infty} \frac{d\Om}{4\pi\sqrt{\Om} \, r}
\left(
a_{\Om} e^{- i\Om U} + a^{\dag}_{\Om} e^{i\Om U}
+ \tilde{a}_{\Om} e^{- i\Om V} + \tilde{a}^{\dag}_{\Om} e^{i\Om V}
\right)
\label{field-a} \\
&= \int_0^{\infty} \frac{d\om}{4\pi\sqrt{\om} \, r}
\left(
b_{\om} e^{- i\om u} + b^{\dag}_{\om} e^{i\om u}
+ \tilde{b}_{\om} e^{- i\om v} + \tilde{b}^{\dag}_{\om} e^{i\om v}
\right).
\label{phi-modes}
\end{align}
The 2D scalar field in the main text differs from this expression
by an overall factor of $r$.
Since $r \simeq a$ is nearly constant in the near-horizon region,
the 2D description is a good approximation.
\hide{
The canonical commutation relations for the creation-annihilation operators include
\begin{align}
&[a_{\Om}, a^{\dag}_{\Om'}] = \d(\Om - \Om'),
\quad
&[\tilde{a}_{\Om}, \tilde{a}^{\dag}_{\Om'}] = \d(\Om - \Om'),
\label{a-CR}
\\
&[b_{\om}, b^{\dag}_{\om'}] = \d(\om - \om'),
\quad
&[\tilde{b}_{\om}, \tilde{b}^{\dag}_{\om'}] = \d(\om - \om').
\label{b-CR}
\end{align}
}
The initial state for the field is typically assumed to be
the Unruh vacuum $| 0 \rangle$ defined by
\be
a_{\Om}|0\rangle = 
\tilde{a}_{\Om}|0\rangle = 0.
\label{vac}
\ee

The Bogoliubov transformation between the creation-annihilation operators 
$(b^{\dag}_{\om}, b_{\om})$ and $(a^{\dag}_{\Om}, a_{\Om})$ is given by
\begin{align}
b_{\om} &= \int_0^{\infty} d\Om \, 
\left(\a_{\om\Om} a_{\Om} + \b_{\om\Om} a^{\dag}_{\Om}\right),
\label{Bogo-1}
\\
b^{\dag}_{\om} &= \int_0^{\infty} d\Om \, 
\left(\b^{\ast}_{\om\Om} a_{\Om} + \a^{\ast}_{\om\Om} a^{\dag}_{\Om}\right),
\label{Bogo-2}
\end{align}
where
\begin{align}
\a_{\om\Om} &\equiv \frac{1}{2\pi} \sqrt{\frac{\om}{\Om}}
\int_{-\infty}^{\infty} du \, e^{i\om u - i\Om U(u)}
\simeq \frac{a}{\pi} \sqrt{\frac{\om}{\Om}} (2a\Om)^{i2a\om}
e^{\pi a\om} \Gamma(-i 2a\om),
\label{alpha}
\\
\b_{\om\Om} &\equiv \frac{1}{2\pi} \sqrt{\frac{\om}{\Om}}
\int_{-\infty}^{\infty} du \, e^{i\om u + i\Om U(u)}
\simeq \frac{a}{\pi} \sqrt{\frac{\om}{\Om}} (2a\Om)^{i2a\om}
e^{- \pi a\om} \Gamma(-i 2a\om).
\label{beta}
\end{align}

Using eqs.\eqref{vac}, \eqref{Bogo-1}, \eqref{Bogo-2}, \eqref{beta}, and \eqref{creation-op}
one can show that the VEV of the number operator
is given by the Planck distribution
\be
\langle 0 | b^{\dag}_{(\om_0, u_0)} b_{(\om_0, u_0)} | 0 \rangle
= \frac{1}{e^{4\pi a\om_0} - 1}.
\ee
This is the standard result of Hawking radiation.


\section{Cutoff Effect in CFT}
\label{Schwarzian}


The energy flux of Hawking radiation can be calculated in 2D CFT
as a Schwarzian derivative relating the conformal transformation
between $U$ 
(the retarded light-cone coordinate around the horizon)
and $u$
(the retarded light-cone coordinate at large distances).
The holomorphic energy-momentum tensor $T_{zz}(z) = T(z)$ 
in a 2D CFT can be defined as
\begin{align}
T(z) = - \frac{1}{2} \lim_{\d \rightarrow 0} \left[
\del_z \phi\left(z+\frac{\d}{2}\right) \del_z \phi\left(z-\frac{\d}{2}\right) + \frac{1}{\d^2}
\right]
\label{Tzz}
\end{align}
via the point-splitting regularization.
Over a conformal map $z \rightarrow w(z)$,
we have
\begin{align}
\del_z \phi(z) \rightarrow \frac{\del w}{\del z} \del_w\phi'(w),
\end{align}
and, as a result,
\begin{align}
T(z) &=
\left(\frac{\del w}{\del z}\right)^2 T'(w) 
+ \frac{1}{2} \lim_{\d \rightarrow 0} \left[
\frac{w^{(1)}(z+\d/2) w^{(1)}(z-\d/2)}{(w(z+\d/2) - w(z-\d/2))^2} - \frac{1}{\d^2}
\right]
\nn \\
&=
\left(\frac{\del w}{\del z}\right)^2 T'(w) 
+ \frac{1}{12}\left[\frac{w^{(3)}}{w^{(1)}} - \frac{3}{2}\left(\frac{w^{(2)}}{w^{(1)}}\right)^2\right],
\label{Tz-Tw}
\end{align}
where $w^{(n)}$ is the $n$-th derivative of $w(z)$ with respect to $z$.
The 2nd term above is called the Schwarzian derivative.

\hide{
Using eq.\eqref{U-def} and
identifying $w$ with $U$ and $z$ with $u$,
the energy flux for Hawking radiation is
\begin{align}
- \frac{1}{12}\left[\frac{w^{(3)}}{w^{(1)}} - \frac{3}{2}\left(\frac{w^{(2)}}{w^{(1)}}\right)^2\right]
= \frac{1}{96 a^2}.
\end{align}
The identity \eqref{Tz-Tw} above suggests a regularized version of this energy flux as
\begin{align}
\frac{1}{2}\left[
\frac{U^{(1)}(u+\d/2) U^{(1)}(u-\d/2)}{(U(u+\d/2) - U(u-\d/2))^2} - \frac{1}{\d^2}
\right]
\end{align}
for a fixed cutoff length $\d$ imposed on the coordinate $u$.
To obtain an approximate Hawking radiation,
we only need $\d \ll a$.

To investigate the effect of a cutoff on the coordinate $U$,
we rewrite the energy flux of Hawking radiation 
by interchanging the roles of $w$ and $z$.
}

Using eq.\eqref{U-def} and
identifying $w = u$ and $z = U$,
the energy flux of Hawking radiation
can be derived as the (scaled) Schwarzian derivative
\begin{align}
\left(\frac{\del w}{\del z}\right)^{-2}
\frac{1}{12}\left[\frac{w^{(3)}}{w^{(1)}} - \frac{3}{2}\left(\frac{w^{(2)}}{w^{(1)}}\right)^2\right]
= \frac{1}{96 a^2}.
\label{HR}
\end{align}

A UV cutoff can be realized as a finite point-splitting parameter $\d$.
With a finite point-splitting regularization
(without taking the limit $\d \rightarrow 0$),
the Hawking radiation energy flux with a UV cutoff is
\begin{align}
&
\left(w^{(1)}(z+\d/2) w^{(1)}(z-\d/2)\right)^{-1}
\left[
\frac{w^{(1)}(z+\d/2) w^{(1)}(z-\d/2)}{(w(z+\d/2) - w(z-\d/2))^2} - \frac{1}{\d^2}
\right]
\nn \\
&\qquad =
\frac{1}{\left[\log\left(\frac{-(U+\d/2)}{-(U-\d/2)}\right)\right]^2} - \frac{(U+\d/2)(U-\d/2)}{\d^2}.
\label{HR-2}
\end{align}
The $\d \rightarrow 0$ limit of eq.\eqref{HR-2} reproduces eq.\eqref{HR}.
But for a given finite $\d > 0$,
the quantity above approaches $0$
as $U \rightarrow - \d/2$.
(The point-splitting regularization is valid only for $U < -\d/2$.)
Hawking radiation diminishes before
the collapsing matter crosses the horizon.
This agrees with our result in Sec.\ref{HR-spectrum}.


\section{Commutation Relation vs Hawking Radiation}
\label{B}


In this section,
we establish a connection between Hawking radiation
and the commutation relation for creation and annihilation operators.

Without assuming the canonical commutation relation 
among $(a^{\dag}_{\Om}, a_{\Om})$, or $(b^{\dag}_{\om}, b_{\om})$,
if we assume that
\be
[a_{\Om}, a_{\Om'}] = [a^{\dag}_{\Om}, a^{\dag}_{\Om'}] \simeq 0,
\label{aa=0}
\ee
(in the sense that they are much smaller than $[a_{\Om}, a^{\dag}_{\Om'}]$),
we can derive from the Bogoliubov transformations \eqref{Bogo-1} and \eqref{Bogo-2} that
\begin{align}
&[b_{(\om_0, u_0)}, b^{\dag}_{(\om_0, u_0)}]
=
\int_0^{\infty} d\om_1 d\om_2 \,
f^{\ast}_{\om_0}(\om_1) f_{\om_0}(\om_2) \,
e^{- i (\om_1 - \om_2) u_0}
\times
\nn \\
&\qquad \times
\int_0^{\infty} d\Om_1 d\Om_2 \,
\left(
\a_{\om_1\Om_1} \a_{\om_2\Om_2}^{\ast}
- \b_{\om_1\Om_2} \b_{\om_2\Om_1}^{\ast}
\right)
[a_{\Om_1}, a^{\dag}_{\Om_2}]
\nn \\
&\simeq
\left(e^{4\pi a \om_0} - 1\right)
\int_0^{\infty} d\om_1 d\om_2 \,
f^{\ast}_{\om_0}(\om_1) f_{\om_0}(\om_2) \,
\int_0^{\infty} d\Om_1 d\Om_2 \,
\b_{\om_1\Om_1} \b_{\om_2\Om_2}^{\ast}
[a_{\Om_1}, a^{\dag}_{\Om_2}],
\label{bb-aa}
\end{align}
where we have used eqs.\eqref{alpha}, \eqref{beta},
and the assumption that
the wave packet $f_{\om_0}$ has a narrow distribution $\Delta \om \ll 1/a$.

Assuming Unruh's vacuum defined as
\be
a_{\Om}|0\rangle = 0,
\label{a0=0}
\ee
and using eqs.\eqref{Number},
we find
\begin{align}
\langle 0 | {\cal N}_{(\om_0, u_0)} | 0 \rangle
&=
\int_0^{\infty} d\om_1 d\om_2 \,
f_{\om_0}(\om_1) f^{\ast}_{\om_0}(\om_2) \,
e^{i (\om_1 - \om_2) u_0}
\int_0^{\infty} d\Om_1 d\Om_2 \,
\b_{\om_1\Om_1}^{\ast} \b_{\om_2\Om_2} 
\langle 0 | a_{\Om_1} a^{\dag}_{\Om_2} | 0 \rangle
\nn \\
&\simeq 
\frac{1}{e^{4\pi a \om_0} - 1} \,
\langle 0 |
[b_{(\om_0, u_0)}, b^{\dag}_{(\om_0, u_0)}]
| 0 \rangle,
\label{N-bb}
\end{align}
where we have used eq.\eqref{bb-aa} in the last step.

Since the commutation relation $[b_{(\om_0, u_0)}, b^{\dag}_{(\om_0, u_0)}]$ is changed,
we should no longer identify $b_{(\om_0, u_0)}$ and $b^{\dag}_{(\om_0, u_0)}$
as the annihilation and creation operators,
or ${\cal N}_{(\om_0, u_0)}$ as the number operator.
Nevertheless,
as long as the operators $a_{\Om}, a^{\dag}_{\Om}, b_{\om}, b^{\dag}_{\om}$
are still defined by the same mode expansions \eqref{field-a} and \eqref{phi-modes} 
of the field $\phi$,
they represent the amplitudes of certain fluctuation modes of $\phi$.
Assuming that an Unruh-DeWit detector of Hawking particles is designed
in the absence of black holes to measure
the magnitude of a fluctuation mode $\psi_{(\om_0, u_0)}$ of the field $\phi$,
the transition probability of the detector measures
the VEV of the operator ${\cal N}_{(\om_0, u_0)}$ defined by eq.\eqref{Number}.
According to this detector, 
the magnitude of Hawking radiation is still $\langle 0 | {\cal N}_{(\om_0, u_0)} | 0 \rangle$.
The conclusion is thus that,
for free-field theories with modified canonical commutation relations
together with the assumptions \eqref{aa=0} and \eqref{a0=0},
the magnitude of Hawking radiation $\langle 0 | {\cal N}_{(\om_0, u_0)} | 0 \rangle$
is simply the Planck distribution times
the commutator $[b_{(\om_0, u_0)}, b^{\dag}_{(\om_0, u_0)}]$.
An $\mathcal{O}(1)$-correction to Hawking radiation implies
an $\mathcal{O}(1)$-correction to the commutator $[b_{(\om_0, u_0)}, b^{\dag}_{(\om_0, u_0)}]$.



\section{Comment on Multi-Particle States}
\label{D}


In this section,
we define the 2-particle state $|\psi, \psi'\rangle$
and derive the transition amplitude for the process 
$|0\rangle\rightarrow|\psi,\psi'\rangle$ \eqref{0-2} 
induced by the interaction \eqref{HV}.

First, 
we define the 2-particle states $|\psi,\psi'\rangle$,
which is chosen to be slightly different from the state
\begin{align}
b^{\dag}_{(\om_0, u_0)} b^{\dag}_{(\om'_0, u'_0)} | 0 \rangle.
\label{bbb}
\end{align}
Eq.\eqref{bbb} is the state obtained by adding two particles 
(for distant observers) on top of Hawking radiation.
It was adopted as the final state in Ref.\cite{Ho:2021sbi}.

\hide{
The reason why we shall choose a different 2-particle state as the final state is the following.
Since the state $b^{\dag}_{(\om_0, u_0)}|0\rangle$ is 
defined by a wave packet $\psi_{(\om_0, u_0)}$ \eqref{wave-packet}
composed of purely positive frequency ($\om > 0$) modes,
its frequency distribution $f_{\om_0}(\om)$
equals the product of the step function $\Th(\om)$ with itself.
The wave packet $\psi_{(\om_0, u_0)}(u)$ is thus 
the convolution of the Fourier transforms of $f_{\om_0}(\om)$ and $\Th(\om)$.
As the Fourier transform of $\Th(\om)$ is $i/(2\pi u)$,
the wave packet satisfies
\begin{align}
\psi_{(\om_0, u_0)} = \frac{i}{\pi} \int_{-\infty}^{\infty} du' \frac{\psi_{(\om_0, u_0)}(u')}{u - u'}
= \frac{i}{\pi} \sum_{k=0}^{\infty}
\frac{1}{u^{k+1}}
\left[
\int_{-\infty}^{\infty} du' \, {u'}^k \psi_{(\om_0, u_0)}(u')
\right]
\label{psi-1/u}
\end{align}
at large $u$.
Generically,
$\left[\int_{-\infty}^{\infty} du' \, {u'}^n \psi_{(\om_0, u_0)}(u)\right]$
is not identically $0$ for all $n = 0, 1, 2, \cdots$,
it implies that $\psi_{(\om_0, u_0)} \propto 1/u^n$ for some finite $n > 0$,
and so an integral of the form
$\int_{-\infty}^{\infty} du e^{mu/2a}\psi_{(\om_0, u_0)}(u)$
diverges for $m > 0$.
Since we expect exponential factors $e^{mu/2a}$ with $m > 0$ in the amplitude.
(The exponential factor $e^{mu/2a}$ is expected;
see eq.\eqref{enhance}),
Hence,
the amplitude diverges unless we consider a better localized wave packet than eq.\eqref{psi-1/u},
and it requires negative frequency modes.
}

To define the final state $|\psi,\psi'\rangle$, 
we start with a generic wave packet
\begin{align}
\psi(u) = 
\int_{-\infty}^{\infty} \frac{d\om}{\sqrt{4\pi|\om|}} \,
f_{\om_0}(\om) e^{-i\om (u - u_0)}.
\label{psipsiprime}
\end{align}
The only difference between this equation \eqref{psipsiprime} and eq.\eqref{wave-packet}
is that it includes negative frequency $(\om < 0)$ modes.
Correspondingly,
we modify $b^{\dag}_{(\om_0, u_0)}$ \eqref{creation-op} to
\begin{align}
b^{\dag}_{\psi} \equiv
\int_0^{\infty} d\om \, (1 - e^{-4\pi\om a}) f_{\om_0}(\om) e^{i\om u_0} b^{\dag}_{\om}
+ \int_{-\infty}^0 d\om \, (e^{4\pi|\om| a} - 1) f_{\om_0}(\om) e^{i\om u_0} b_{-\om},
\label{D3}
\end{align}
such that the corresponding 1-particle state
\footnote{
The normalization condition for $\psi$ is
\begin{align}
\langle \psi | \psi \rangle =
\int_0^{\infty} d\om \left[
(1-e^{-4\pi\om a})\left|f_{\om_0}(\om)\right|^2 
+ (e^{4\pi\om a} - 1)\left|f_{\om_0}(-\om)\right|^2
\right].
\end{align}
}
\begin{align}
|\psi\rangle = b^{\dag}_{\psi} | 0 \rangle
\end{align}
satisfies
\begin{align}
\langle 0 | \phi(u) |\psi\rangle = \psi(u).
\end{align}

Similarly,
a 2-particle state defined as
\begin{align}
| \psi, \psi' \rangle
\equiv
\left(b^{\dag}_{\psi} b^{\dag}_{\psi'} - \langle 0|b^{\dag}_{\psi} b^{\dag}_{\psi'}|0\rangle
\right)|0\rangle
\label{2psi}
\end{align}
satisfies eq.\eqref{0phiphipsipsi-0}:
\begin{align}
\langle 0 | \phi(u)\phi(u') | \psi, \psi' \rangle
= \psi(u)\psi'(u) + \psi(u')\psi'(u).
\label{0phiphipsipsi}
\end{align}

Using eqs.\eqref{HV} and \eqref{0phiphipsipsi},
we have
\begin{align}
\langle \psi, \psi' | H_k(V') | 0 \rangle
\simeq 
- \frac{\lam_k}{M_p^{4k-2}} \int
dU
(R_{VV}(U, V))^k (\del_U^k\psi^{\ast}) (\del_U^k\psi^{\prime\ast}).
\end{align}
Suppose the product
$(\del_U^k\psi^{\ast}) (\del_U^k\psi^{\prime\ast})$ has a finite support of $\Delta u$,
its time span in terms of the coordinate $U$ is
\be
\Delta U \sim \Delta u \, e^{-u/2a},
\ee
which is approximately $0$ for $u \gg \mathcal{O}(a)$.
Since we will be interested in the scrambling time $u \sim \mathcal{O}(a\log(aM_p))$,
it is a good approximation to write
\be
(R_{VV}(U, V))^k (\del_U^k\psi^{\ast}) (\del_U^k\psi^{\prime\ast})
\simeq
(R_{VV}(0, V))^k (\del_U^k\psi^{\ast}) (\del_U^k\psi^{\prime\ast}).
\ee
That is,
since the outgoing wave packets $\psi, \psi'$ are
originated from the near-horizon region at late times (large $u$),
only the background field in the near-horizon region ($U\simeq 0$) is relevant
to the scattering amplitude.
This also explains why eq.\eqref{HV} is a good approximation,
as we commented in the paragraph below eq.\eqref{HV}.

We can thus rewrite the transition amplitude \eqref{Ak-0} as
\begin{align}
{\cal A}_k(V)
&\simeq
i \frac{\lam_k}{M_p^{4k-2}} \,
\left[\int_{V_0}^{V} dV' \, (R_{VV}(0, V'))^k\right]
K_k,
\label{A-K}
\end{align}
where
\begin{align}
K_k
&\equiv
\int_{-\infty}^0 dU \, \del_U^k\psi^{\ast}(u) \del_U^k\psi^{\prime\ast}(u)
\nn \\
&=
\int_{-\infty}^{\infty} du \, e^{(2k-1)\frac{u}{2a}}
\left[\hat{D}_u(k) \psi^{\ast}(u)\right] \left[\hat{D}_u(k) \psi^{\prime\ast}(u)\right],
\label{K-k-0}
\end{align}
and
\begin{align}
\hat{D}_u(k) 
\equiv e^{-k\frac{u}{2a}} \left(e^{\frac{u}{2a}} \del_u\right)^k.
\end{align}
Since the wave packet $\psi_{(\om_0, u_0)}(u)$ is defined in terms of $u$,
they are restricted to the region $U \in (-\infty, 0)$.
We can thus replace the integral $\int_{-\infty}^{\infty} dU$
by $\int_{-\infty}^0 dU$ as we did in eq.\eqref{K-k-0}.

For example,
for the Gaussian wave packets
\begin{align}
\psi^{\ast}(u) &=
\frac{1}{\sqrt{2\sqrt{\pi} \om_0\Delta u}} \, e^{-\frac{(u-u_0)^2}{2\Delta u^2} + i\om_0(u - u_0)},
\\
\psi^{\prime\ast}(u) &=
\frac{1}{\sqrt{2\sqrt{\pi} \om_0\Delta u}} \, e^{-\frac{(u-u'_0)^2}{2\Delta u^2} + i\om_0(u - u'_0)}.
\label{Gaussian-wave}
\end{align}
with $\om_0 \Delta u \gg 1$,
we have
\begin{align}
\hat{D}_u(k) \psi^{\ast}(u)
\simeq 
D(k) \psi^{\ast}(u),
\end{align}
and similarly for $\psi'$,
where $D(k)$ is defined by eq.\eqref{D(k)-1}.
It is then straightforward to evaluate $K_k$ \eqref{K-k-0} as
\begin{align}
K_k
\hide{
&\simeq
\frac{D^2(k)}{2\sqrt{\pi} \om_0\Delta u} \, 
\int_{-\infty}^{\infty} du \, e^{(2k-1)\frac{u}{2a}} 
e^{-\frac{(u-u_0)^2}{2\Delta u^2} + i\om_0 (u-u_0)}
e^{-\frac{(u-u'_0)^2}{2\Delta u^2} + i\om_0 (u'-u_0)}
\nn \\
}
&=
\frac{D^2(k)}{2 \om_0} \, 
e^{\frac{(2k-1)}{2a}\left(\frac{u_0+u'_0}{2}\right)}
e^{-\frac{(u_0 - u'_0)^2}{4\Delta u^2}}
e^{-\om_0^2\Delta u^2} 
e^{\frac{(2k-1)^2}{(4a)^2}\Delta u^2} 
e^{i(2k-1)\om_0\Delta u^2/2a}.
\label{K-k-1}
\end{align}
Plugging it back into eq.\eqref{A-K},
and taking the absolute value,
we obtain the magnitude of the amplitude \eqref{Ak-result} as
\begin{align}
|{\cal A}_k(V)|
&\equiv
\frac{|\lam_k|}{M_p^{4k-2}} \,
\left|\int_{V_0}^{V} dV' \, (R_{VV}(0, V'))^k\right|
\frac{|D^2(k)|}{2 \om_0} \, 
e^{\frac{(2k-1)}{2a}\left(\frac{u_0+u'_0}{2}\right)}
e^{-\frac{(u_0 - u'_0)^2}{4\Delta u^2}}
e^{-\om_0^2\Delta u^2} 
e^{\frac{(2k-1)^2}{(4a)^2}\Delta u^2}.
\label{A-K-result-0}
\end{align}



\vskip .8cm
\baselineskip 22pt
\begin{thebibliography}{99}
\itemsep 0pt







\bibitem{trans-Planckian-1}
  G.~'t Hooft,
  ``On the Quantum Structure of a Black Hole,''
  Nucl.\ Phys.\ B {\bf 256}, 727 (1985).
  doi:10.1016/0550-3213(85)90418-3

\bibitem{trans-Planckian-4}
  A.~D.~Helfer,
  ``Do black holes radiate?,''
  Rept.\ Prog.\ Phys.\  {\bf 66}, 943 (2003)
  doi:10.1088/0034-4885/66/6/202
  [gr-qc/0304042].

\bibitem{Gryb:2018pur}
S.~Gryb, P.~Palacios and K.~Th\'ebault,
``On the Universality of Hawking Radiation,''
Brit. J. Phil. Sci. \textbf{78}, 809-837 (2021)
doi:10.1093/bjps/axz025
[arXiv:1812.07078 [physics.hist-ph]].

\bibitem{Polchinski:1995ta}
J.~Polchinski,
``String theory and black hole complementarity,''
[arXiv:hep-th/9507094 [hep-th]].

\bibitem{Giddings}
S.~B.~Giddings,
``Schr\"odinger evolution of the Hawking state,''
Phys. Rev. D \textbf{102}, 125022 (2020)
doi:10.1103/PhysRevD.102.125022
[arXiv:2006.10834 [hep-th]];
S.~B.~Giddings,
``Schr\"odinger evolution of two-dimensional black holes,''
JHEP \textbf{12}, 025 (2021)
doi:10.1007/JHEP12(2021)025
[arXiv:2108.07824 [hep-th]];
S.~B.~Giddings and J.~Perkins,
``Quantum evolution of the Hawking state for black holes,''
[arXiv:2204.13126 [hep-th]].

\bibitem{Unruh:1994je} 
  W.~G.~Unruh,
  ``Sonic analog of black holes and the effects of high frequencies on black hole evaporation,''
  Phys.\ Rev.\ D {\bf 51}, 2827 (1995).
  doi:10.1103/PhysRevD.51.2827

\bibitem{Brout:1995wp} 
  R.~Brout, S.~Massar, R.~Parentani and P.~Spindel,
  ``Hawking radiation without transPlanckian frequencies,''
  Phys.\ Rev.\ D {\bf 52}, 4559 (1995)
  doi:10.1103/PhysRevD.52.4559
  [hep-th/9506121].

\bibitem{Corley:1996ar}
S.~Corley and T.~Jacobson,
``Hawking spectrum and high frequency dispersion,''
Phys. Rev. D \textbf{54}, 1568-1586 (1996)
doi:10.1103/PhysRevD.54.1568
[arXiv:hep-th/9601073 [hep-th]].

\bibitem{Barcelo:2005fc}
C.~Barcelo, S.~Liberati and M.~Visser,
``Analogue gravity,''
Living Rev. Rel. \textbf{8}, 12 (2005)
doi:10.12942/lrr-2005-12
[arXiv:gr-qc/0505065 [gr-qc]].

\bibitem{Hambli:1995pp}
N.~Hambli and C.~P.~Burgess,
``Hawking radiation and ultraviolet regulators,''
Phys. Rev. D \textbf{53}, 5717-5722 (1996)
doi:10.1103/PhysRevD.53.5717
[arXiv:hep-th/9510159 [hep-th]].

\bibitem{Lubo:2003rs}
M.~Lubo,
``Ultraviolet cutoff, black hole radiation equilibrium and big bang,''
Phys. Rev. D \textbf{68}, 125005 (2003)
doi:10.1103/PhysRevD.68.125005
[arXiv:hep-th/0306187 [hep-th]].


\bibitem{Unruh:2004zk}
W.~G.~Unruh and R.~Schutzhold,
``On the universality of the Hawking effect,''
Phys. Rev. D \textbf{71}, 024028 (2005)
doi:10.1103/PhysRevD.71.024028
[arXiv:gr-qc/0408009 [gr-qc]].

\bibitem{Agullo:2009wt}
I.~Agullo, J.~Navarro-Salas, G.~J.~Olmo and L.~Parker,
``Insensitivity of Hawking radiation to an invariant Planck-scale cutoff,''
Phys. Rev. D \textbf{80}, 047503 (2009)
doi:10.1103/PhysRevD.80.047503
[arXiv:0906.5315 [gr-qc]].

\bibitem{Kajuri:2018myh}
N.~Kajuri and D.~Kothawala,
``Universality of Hawking radiation in non local field theories,''
Phys. Lett. B \textbf{791}, 319-322 (2019)
doi:10.1016/j.physletb.2019.03.006
[arXiv:1806.10345 [gr-qc]].

\bibitem{Ho:2021sbi}
P.~M.~Ho, H.~Kawai and Y.~Yokokura,
``Planckian physics comes into play at Planckian distance from horizon,''
JHEP \textbf{01}, 019 (2022)
doi:10.1007/JHEP01(2022)019
[arXiv:2111.01967 [hep-th]].

\bibitem{HR}
M.~Visser,
``Essential and inessential features of Hawking radiation,''
Int. J. Mod. Phys. D \textbf{12}, 649-661 (2003)
doi:10.1142/S0218271803003190
[arXiv:hep-th/0106111 [hep-th]].
C.~Barcelo, S.~Liberati, S.~Sonego and M.~Visser,
``Quasi-particle creation by analogue black holes,''
Class. Quant. Grav. \textbf{23}, 5341-5366 (2006)
doi:10.1088/0264-9381/23/17/014
[arXiv:gr-qc/0604058 [gr-qc]].
C.~Barcelo, S.~Liberati, S.~Sonego and M.~Visser,
``Hawking-like radiation does not require a trapped region,''
Phys. Rev. Lett. \textbf{97}, 171301 (2006)
doi:10.1103/PhysRevLett.97.171301
[arXiv:gr-qc/0607008 [gr-qc]].


\bibitem{Leahy:1983vb}
D.~A.~Leahy and W.~G.~Unruh,
``EFFECTS OF A LAMBDA PHI**4 INTERACTION ON BLACK HOLE EVAPORATION IN TWO-DIMENSIONS,''
Phys. Rev. D \textbf{28}, 694-702 (1983)
doi:10.1103/PhysRevD.28.694

\bibitem{Akhmedov:2015xwa}
E.~T.~Akhmedov, H.~Godazgar and F.~K.~Popov,
``Hawking radiation and secularly growing loop corrections,''
Phys. Rev. D \textbf{93}, no.2, 024029 (2016)
doi:10.1103/PhysRevD.93.024029
[arXiv:1508.07500 [hep-th]].

\bibitem{Sekino:2008he}
Y.~Sekino and L.~Susskind,
``Fast Scramblers,''
JHEP \textbf{10}, 065 (2008)
doi:10.1088/1126-6708/2008/10/065
[arXiv:0808.2096 [hep-th]].

\bibitem{Hawking:1974sw} 
  S.~W.~Hawking,
  ``Particle Creation by Black Holes,''
  Commun.\ Math.\ Phys.\  {\bf 43}, 199 (1975)
  [Commun.\ Math.\ Phys.\  {\bf 46}, 206 (1976)].

\bibitem{Amati:1988tn}
D.~Amati, M.~Ciafaloni and G.~Veneziano,
``Can Space-Time Be Probed Below the String Size?,''
Phys. Lett. B \textbf{216}, 41-47 (1989)
doi:10.1016/0370-2693(89)91366-X

\bibitem{Scardigli:1999jh}
F.~Scardigli,
``Generalized uncertainty principle in quantum gravity from micro - black hole Gedanken experiment,''
Phys. Lett. B \textbf{452}, 39-44 (1999)
doi:10.1016/S0370-2693(99)00167-7
[arXiv:hep-th/9904025 [hep-th]].

\bibitem{Garay:1994en}
L.~J.~Garay,
``Quantum gravity and minimum length,''
Int. J. Mod. Phys. A \textbf{10}, 145-166 (1995)
doi:10.1142/S0217751X95000085
[arXiv:gr-qc/9403008 [gr-qc]].

\bibitem{Kempf:1994su}
A.~Kempf, G.~Mangano and R.~B.~Mann,
``Hilbert space representation of the minimal length uncertainty relation,''
Phys. Rev. D \textbf{52}, 1108-1118 (1995)
doi:10.1103/PhysRevD.52.1108
[arXiv:hep-th/9412167 [hep-th]].

\bibitem{Kempf:1996fz}
A.~Kempf,
``Nonpointlike particles in harmonic oscillators,''
J. Phys. A \textbf{30}, 2093-2102 (1997)
doi:10.1088/0305-4470/30/6/030
[arXiv:hep-th/9604045 [hep-th]].

\bibitem{Brau:1999uv}
F.~Brau,
``Minimal length uncertainty relation and hydrogen atom,''
J. Phys. A \textbf{32}, 7691-7696 (1999)
doi:10.1088/0305-4470/32/44/308
[arXiv:quant-ph/9905033 [quant-ph]].

\bibitem{Maggiore:1993rv}
M.~Maggiore,
``A Generalized uncertainty principle in quantum gravity,''
Phys. Lett. B \textbf{304}, 65-69 (1993)
doi:10.1016/0370-2693(93)91401-8
[arXiv:hep-th/9301067 [hep-th]].

\bibitem{Hossenfelder:2012jw}
S.~Hossenfelder,
``Minimal Length Scale Scenarios for Quantum Gravity,''
Living Rev. Rel. \textbf{16}, 2 (2013)
doi:10.12942/lrr-2013-2
[arXiv:1203.6191 [gr-qc]].

\bibitem{Snyder:1946qz}
H.~S.~Snyder,
``Quantized space-time,''
Phys. Rev. \textbf{71}, 38-41 (1947)
doi:10.1103/PhysRev.71.38

\bibitem{Chu:1998qz}
C.~S.~Chu and P.~M.~Ho,
``Noncommutative open string and D-brane,''
Nucl. Phys. B \textbf{550}, 151-168 (1999)
doi:10.1016/S0550-3213(99)00199-6
[arXiv:hep-th/9812219 [hep-th]].

\bibitem{Seiberg:1999vs}
N.~Seiberg and E.~Witten,
``String theory and noncommutative geometry,''
JHEP \textbf{09}, 032 (1999)
doi:10.1088/1126-6708/1999/09/032
[arXiv:hep-th/9908142 [hep-th]].

\bibitem{Amelino-Camelia:2000stu}
G.~Amelino-Camelia,
``Relativity in space-times with short distance structure governed by an observer independent (Planckian) length scale,''
Int. J. Mod. Phys. D \textbf{11}, 35-60 (2002)
doi:10.1142/S0218271802001330
[arXiv:gr-qc/0012051 [gr-qc]].

\bibitem{Amelino-Camelia:2000cpa}
G.~Amelino-Camelia,
``Testable scenario for relativity with minimum length,''
Phys. Lett. B \textbf{510}, 255-263 (2001)
doi:10.1016/S0370-2693(01)00506-8
[arXiv:hep-th/0012238 [hep-th]].

\bibitem{Yoneya:1987gb}
T.~Yoneya,
``DUALITY AND INDETERMINACY PRINCIPLE IN STRING THEORY,''
in “Wandering in the Fields”,
eds. K. Kawarabayashi and A. Ukawa (World Scientific, 1987), p. 419;
see also
String Theory and Quantum Gravity in “Quantum String Theory”,
eds. N. Kawamoto and T. Kugo (Springer, 1988), p. 23.
UT-KOMABA-87-3.

\bibitem{Yoneya:1989ai}
T.~Yoneya,
``On the Interpretation of Minimal Length in String Theories,''
Mod. Phys. Lett. A \textbf{4}, 1587 (1989)
doi:10.1142/S0217732389001817

\bibitem{Yoneya:2000bt}
T.~Yoneya,
``String theory and space-time uncertainty principle,''
Prog. Theor. Phys. \textbf{103}, 1081-1125 (2000)
doi:10.1143/PTP.103.1081
[arXiv:hep-th/0004074 [hep-th]].

\bibitem{Chu:1999wz}
C.~S.~Chu, P.~M.~Ho and Y.~C.~Kao,
``World volume uncertainty relations for D-branes,''
Phys. Rev. D \textbf{60}, 126003 (1999)
doi:10.1103/PhysRevD.60.126003
[arXiv:hep-th/9904133 [hep-th]].

\bibitem{Brandenberger:2002nq}
R.~Brandenberger and P.~M.~Ho,
``Noncommutative space-time, stringy space-time uncertainty principle, and density fluctuations,''
Phys. Rev. D \textbf{66}, 023517 (2002)
doi:10.1103/PhysRevD.66.023517
[arXiv:hep-th/0203119 [hep-th]].

\bibitem{Matusis:2000jf}
A.~Matusis, L.~Susskind and N.~Toumbas,
``The IR / UV connection in the noncommutative gauge theories,''
JHEP \textbf{12}, 002 (2000)
doi:10.1088/1126-6708/2000/12/002
[arXiv:hep-th/0002075 [hep-th]].

\bibitem{YF}
C.~N. Yang and D.~Feldman,
``The S-Matrix in the Heisenberg Representation,''
Phys.\ Rev.\ {\bf 79}, 972 (1950).

\bibitem{Cheng:2001du}
T.~C.~Cheng, P.~M.~Ho and M.~C.~Yeh,
``Perturbative approach to higher derivative and nonlocal theories,''
Nucl. Phys. B \textbf{625}, 151-165 (2002)
doi:10.1016/S0550-3213(02)00020-2
[arXiv:hep-th/0111160 [hep-th]].
T.~C.~Cheng, P.~M.~Ho and M.~C.~Yeh,
``Perturbative approach to higher derivative theories with fermions,''
Phys. Rev. D \textbf{66}, 085015 (2002)
doi:10.1103/PhysRevD.66.085015
[arXiv:hep-th/0206077 [hep-th]].

\bibitem{Ho:2020cbf}
P.~M.~Ho and Y.~Yokokura,
``Firewall From Effective Field Theory,''
[arXiv:2004.04956 [hep-th]].

\bibitem{Ho:2020cvn}
P.~M.~Ho,
``From Uneventful Horizon to Firewall in D-Dimensional Effective Theory,''
[arXiv:2005.03817 [hep-th]].

\bibitem{Iso:2008sq}
S.~Iso,
``Hawking Radiation, Gravitational Anomaly and Conformal Symmetry: The Origin of Universality,''
Int. J. Mod. Phys. A \textbf{23}, 2082-2090 (2008)
doi:10.1142/S0217751X08040627
[arXiv:0804.0652 [hep-th]].

\bibitem{Davies:1976ei}
P.~Davies, S.~Fulling and W.~Unruh,
``Energy Momentum Tensor Near an Evaporating Black Hole,''
Phys.\ Rev.\ D \textbf{13}, 2720-2723 (1976)
doi:10.1103/PhysRevD.13.2720

\bibitem{Brout:1995rd} 
  R.~Brout, S.~Massar, R.~Parentani and P.~Spindel,
  ``A Primer for black hole quantum physics,''
  Phys.\ Rept.\  {\bf 260}, 329 (1995)
  doi:10.1016/0370-1573(95)00008-5
  [arXiv:0710.4345 [gr-qc]].

\bibitem{Ho:2019pjr}
P.~M.~Ho, Y.~Matsuo and Y.~Yokokura,
``Analytic description of semiclassical black-hole geometry,''
Phys. Rev. D \textbf{102}, no.2, 024090 (2020)
doi:10.1103/PhysRevD.102.024090
[arXiv:1912.12855 [hep-th]].

\bibitem{Ho:2019qiu}
P.~M.~Ho, Y.~Matsuo and Y.~Yokokura,
``Distance between collapsing matter and apparent horizon in evaporating black holes,''
Phys. Rev. D \textbf{104}, no.6, 064005 (2021)
doi:10.1103/PhysRevD.104.064005
[arXiv:1912.12863 [gr-qc]].

\bibitem{Kawai:2013mda} 
  H.~Kawai, Y.~Matsuo and Y.~Yokokura,
  ``A Self-consistent Model of the Black Hole Evaporation,''
  Int.\ J.\ Mod.\ Phys.\ A {\bf 28}, 1350050 (2013)
  [arXiv:1302.4733 [hep-th]].

\bibitem{Kawai:2014afa} 
  H.~Kawai and Y.~Yokokura,
  ``Phenomenological Description of the Interior of the Schwarzschild Black Hole,''
  Int.\ J.\ Mod.\ Phys.\ A {\bf 30}, 1550091 (2015)
  doi:10.1142/S0217751X15500918
  [arXiv:1409.5784 [hep-th]].
  H.~Kawai and Y.~Yokokura,
  ``A Model of Black Hole Evaporation and 4D Weyl Anomaly,''
  Universe {\bf 3}, no. 2, 51 (2017)
  doi:10.3390/universe3020051
  [arXiv:1701.03455 [hep-th]].

\bibitem{Kawai:2020rmt}
H.~Kawai and Y.~Yokokura,
``Black Hole as a Quantum Field Configuration,''
Universe \textbf{6}, no.6, 77 (2020)
doi:10.3390/universe6060077
[arXiv:2002.10331 [hep-th]].

\bibitem{Kawai:2021qdk}
H.~Kawai and Y.~Yokokura,
``Interior metric of slowly formed black holes in a heat bath,''
Phys. Rev. D \textbf{105}, no.4, 045017 (2022)
doi:10.1103/PhysRevD.105.045017
[arXiv:2108.02242 [hep-th]].














\end{thebibliography}
\end{document}